\documentclass[twocolumn]{aastex62}

\hypersetup{linkcolor=red,citecolor=green,filecolor=cyan,urlcolor=magenta}

\graphicspath{{./}{figures/}}
\usepackage{amsmath}
\shorttitle{NICMOS KPI II: Demographics of BDs}
\shortauthors{Factor \& Kraus}

\begin{document}

\title{NICMOS Kernel-Phase Interferometry II: \\Demographics of Nearby Brown Dwarfs}

\correspondingauthor{Samuel M. Factor}
\email{sfactor@utexas.edu}

\author[0000-0002-8332-8516]{Samuel M. Factor}
\affil{Department of Astronomy, The University of Texas at Austin, Austin, TX 78712, USA}

\author[0000-0001-9811-568X]{Adam L. Kraus}
\affil{Department of Astronomy, The University of Texas at Austin, Austin, TX 78712, USA}

\begin{abstract}
%250 word limit

Star formation theories have struggled to reproduce binary brown dwarf population demographics (frequency, separation, mass-ratio). Kernel-phase interferometry is sensitive to companions at separations inaccessible to classical imaging, enabling tests of formation at new physical scales below the hydrogen burning limit. 
We analyze the detections and sensitivity limits from our previous kernel-phase analysis of archival HST/NICMOS surveys of field brown dwarfs. After estimating physical properties of the 105 late M to T dwarfs using Gaia distances and evolutionary models, we use a Bayesian framework to compare these results to a model companion population defined by log-normal separation and power-law mass-ratio distributions. 
When correcting for Malmquist bias, we find a companion fraction of $F=0.11^{+0.04}_{-0.03}$ and a separation distribution centered at $\rho=2.2^{+1.2}_{-1.0}$~au, smaller and tighter than seen in previous studies. We also find a mass-ratio power-law index which strongly favors equal-mass systems: $\gamma=4.0^{+1.7}_{-1.5}-11^{+4}_{-3}$ depending on the assumed age of the field population ($0.9-3.1$~Gyr). We attribute the change in values to our use of kernel-phase interferometry which enables us to resolve the peak of the semimajor axis distribution with significant sensitivity to low-mass companions. We confirm the previously-seen trends of decreasing binary fraction with decreasing mass and a strong preference for tight and equal-mass systems in the field-age sub-stellar regime; only $0.9^{+1.1}_{-0.6}$\% of systems are wider than 20~au and $<1.0^{+1.4}_{-0.6}$\% of systems have a mass-ratio $q<0.6$. We attribute this to turbulent fragmentation setting the initial conditions followed by a brief period of dynamical evolution, removing the widest and lowest-mass companions, before the birth cluster dissolves.
\end{abstract}

\section{Introduction} \label{sec:intro}

Star formation, and more specifically binary (or multiple) formation, is a foundational process in astrophysics, contributing to stellar populations and the stellar content of galaxies, interacting binaries and the transients they produce, as well as planet formation and habitability. A successful theory should replicate trends in not only the single star IMF, but also companion mass-ratio and separation distributions as well as frequency as a function of host mass, age, and other fundamental parameters \citep{Duchene2013}. Over the past two decades thousands of brown dwarfs (BDs) have been discovered \citep{UltraCoolSheet2020}, spurring detailed studies of our local neighborhood \citep{Kirkpatrick2019,Best2021} including binary brown dwarfs \citep{Burgasser2007} and dynamical masses \citep{Dupuy2017}, and enabling studies of demographic trends below the hydrogen burning limit, challenging formation theories at new scales \citep{Luhman2012,Offner2022}. 

While no comprehensive theory for multiple star (or BD) formation currently exists, many processes likely play a role, including core fragmentation, disk fragmentation, and dynamical interaction/evolution. Since BDs are significantly below the Jeans mass of a typical core ($\sim1~M_\odot$), some more complicated physics beyond classical collapse must be having an effect. Turbulent fragmentation provides a clear formation pathway for lower mass objects and high-order multiple systems \citep{Bate2002,Bonnell2008,Bate2009,Bate2012,Offner2010,Guszejnov2017}. Disk fragmentation likely plays a minimal roll in forming a BD in the disk around another BD since the mass budget is too low \citep{Burgasser2007}, though it is possible for BD-BD binaries to form as a higher order multiple inside the disk around a more massive star and survive ejection \citep{Stamatellos2009}. 

High-order systems tend to be dynamically unstable, especially when considering the larger star forming surroundings. They evolve on relatively short timescales, ejecting some objects and binding the remnants into tight binary systems \citep{Reipurth2001}. Current state of the art simulations are approaching the spatial scales needed to model brown dwarf binaries and their circumsubstellar disks. Sink particles are seeded and accretion takes place on scales of 0.5--5~au \citep{Bate2003,Bate2009,Bate2012} or even larger \citep{Offner2009,Grudic2022}, on the same scales as (or larger than) a typical brown dwarf binary, while disks are only resolved down to scales of 1--10~au. Historically, a softened Newtonian potential has also been used close to sink particles, enhancing binary disruption. Even so, recent simulations have roughly reproduced the observed BD binary fraction and trends in mass ratio and separation seen in the field population \citep{Luhman2012,Offner2022}, though still struggle to form the somewhat rare widely separated pairs \citep{Radigan2009,Kraus2011,Faherty2020}.

Previous demographic surveys have found a BD binary population heavily skewed toward equal mass-ratios, and modeled the population with a power-law with index $\gamma\sim2-5$ \citep[][surveys which were sensitive to mass-ratios of $q\gtrsim0.2$]{Reid2006,Burgasser2007,Allen2007}. \citet{Fontanive2018} studied later spectral type objects (T5--Y0) and found a slightly stronger mass-ratio power-law index of $\gamma\sim6$ (with sensitivity down to $q\gtrsim0.4$). Direct imaging surveys and searches for co-moving sources in astrometric surveys have constrained the overall BD binary fraction to $\sim20\%$ \citep{Reid2001,Close2002,Burgasser2003b,Bouy2003,Close2003,Gizis2003a,Reid2006,Allen2007,Burgasser2007,Aberasturi2014} with few companions on wide orbits with semimajor axes $\gtrsim20$~au. Surveys using the radial velocity technique, though limited to the brightest targets, have searched for extremely tight companions and similarly found a much lower frequency for companions on orbits with semimajor axes $\lesssim1$~au of $2.5^{+8.6}_{-1.6}\%$ \citep{Blake2010} \citep[see also][]{Basri2006,Joergens2008,Hsu2021}. The semimajor axis distribution has been modeled as log-normal, centered around $\sim6$~au with a width of $0.2-0.3$~dex \citep{Reid2006,Burgasser2007,Allen2007}. The \citet{Fontanive2018} study of later spectral type objects found a roughly similar semimajor axis distribution with a lower overall companion frequency of $8\pm6\%$ (and $2\pm2\%$ for tight companions with separations $<1$~au). In contrast, stellar-mass binaries show a much flatter mass-ratio distribution ($\gamma\sim-2$ to $\sim1$ for A0--M4 primaries), a higher companion fraction ($\sim70\%$ to $\sim35\%$), and a broader semimajor axis distribution centered at wider separations ($\sim390$ to $\sim10$~au) \citep[][see further discussion in Section~\ref{sec:discPrev} and Figure~\ref{fig:paramMass}]{DeRosa2014,Kraus2012,Winters2019}. Yet there is still some uncertainty in the mean of the separation distribution of BD binaries as RV surveys have not run long enough to detect the most common companions and the mean separation appears to be roughly at the inner working angle of direct imaging surveys, which can not yet resolve the closest companions. 

In this work, we take advantage of interferometric analysis to push our inner working angle to tighter separations than previously accessible via classical imaging techniques/analysis and work to resolve the peak of the underlying semimajor axis. We perform a demographic analysis on the catalogue of BD binaries presented in \citet{Factor2022}, built by applying a new kernel-phase interferometry pipeline to the entire \emph{HST/NICMOS} imaging archive of nearby brown dwarfs in F110W and F170M (observed in 7 programs, 3 of which are analyzed in this work and are outlined in Section~\ref{sec:obs}). That work built on previous analysis by searching for companions at tighter separations than was possible with classical PSF subtraction and revisiting candidate companions proposed in a previous kernel-phase analysis \citep{Pope2013}. While no new companions were discovered around these well studied targets, we did confirm one candidate companion and marginally recover a second, both proposed by \citet{Pope2013}, and did not recover other candidates. We also measured detection limits for each target, important for the survey analysis done in this work. In this work, we derive physical parameters from the observed parameters using the method described in Section~\ref{sec:phys} and model the demographic parameters as described in Section~\ref{sec:binPop}. We then compare our results (Section~\ref{sec:res}) to previous surveys and theories of binary formation (Section~\ref{sec:disc}).

\section{Observations} \label{sec:obs}
\subsection{NICMOS Data}\label{sec:nicmos}
We adopt our sample from the catalogue presented in Paper 1 of this series \citep{Factor2022}. That work analyzed archival \emph{HST} observations of field brown dwarfs using the Near Infrared Camera and Multi-Object Spectrometer (NICMOS). Camera 1 of NICMOS has a $256\times256$ pixel detector, with a pixel scale of 43~mas, for a field of view of $11\arcsec~\times~11\arcsec$. Our previous KPI analysis had an outer working angle of $0\farcs5$ due to strong aliasing in the Fourier domain caused by wide separation companions. In this work we supplement our sensitivity to wide companions based on the sensitivity of classical imaging surveys, further described in Section~\ref{sec:binPop}.

A detailed description of the datasets is presented in \citet{Factor2022} but to briefly summarize, that work analyzed data from 7 programs observing brown dwarfs in F110W and F170M (roughly corresponding to \emph{J} and \emph{H} bands). For this work we narrow that sample down to three programs---9833, 10143, and 10879---which conducted an unbiased search for binaries instead of targeting known pairs. Details of those programs are presented in Table~\ref{tab:obs}.

\begin{deluxetable*}{clcccll}
\tabletypesize{\scriptsize}
\tablecaption{Observations\label{tab:obs}}
\tablehead{\colhead{Program ID} & \colhead{P.I.} & \colhead{Cycle} & \colhead{$N_{obj}$} & \colhead{$N_{dithers}$} & \colhead{aprox. epoc} & \colhead{Publication}}
\startdata
9833 & \citet{Burgasser2003c} & 12 & 22 & 3--6 & 9/2003--7/2004 & \citet{Burgasser2006} \\
10143 & \citet{Reid2004} & 13 & 56 & 2 & 9/2004--6/2006 & \citet{Reid2006} \\
10879 & \citet{Reid2006a} & 15 & 28 & 2 & 7/2006--5/2007 & \citet{Reid2008} \\
\enddata
\tablecomments{Program 9833 observed in F090M in addition to F110W and F170M}
\end{deluxetable*}

Table~2 in \citet{Factor2022} details the properties of the targets analyzed in that work, a subset of which are analyzed here. This sample covers 105 targets (including 15 binaries) with spectral types of late M ($\sim10\%$), L ($\sim70\%$), and T ($\sim20\%$) dwarfs (roughly 95--30~$M_\mathrm{Jup}$ depending on the assumed age of the field population) at distances ranging from $\sim$5--35 pc. Analysis of the full sample, presented in \citet{Factor2022}, applied a new KPI pipeline \citep{argus} to these observations using a novel multi-calibrator approach.

Table~4--6 and Figure Sets~5--8 in \citet{Factor2022} present the detection limits for each target analyzed in this work. The method used to derive these limits is described in detail in Section~3.3 of \citet{Factor2022} but to briefly summarize, we scramble the indices of the model subtracted kernel phases (i.e. randomly reorder the residual phase noise) in order to generate a new instance of the intrinsic noise. We then fit the noise on a grid of PA and separation to see what spurious companions are mimicked by noise. In the best cases, significant detections of companions can be achieved up to a contrast of $\sim100:1$ down to a separation of $\sim0\farcs1$ and significantly closer at lower contrast. 

Table~\ref{tab:binPropInt} presents the observational properties of the relevant binary systems while Table~\ref{tab:binPropFin} presents the physical properties derived using the methods described below. A companion is considered a significant detection if it is significantly ($>5\sigma$ confidence) detected in at least 4 calibrators and fitted parameters are consistent with each other within $1\sigma$ in contrast and $5\sigma$ in position. Our sample does not include the marginal detection (2M 2028+0052) but does include the two wide separation companions (2M 0915+0422 and 2M 1707-0558) which were aliased in our earlier KPI analysis but are easily seen in the images. We adopt astrometry and photometry from \citet{Pope2013}.

\section{Methods} \label{sec:meth}

\subsection{Physical Properties of Binaries}\label{sec:phys}
To perform a demographic analysis of detections and detection limits for binary companions we must first convert observed quantities (angular separation and contrast) into physical quantities (projected separation and mass ratio). Distances derived from Gaia eDR3 parallaxes were used \citep{BailerJones2021,Gaia2021} when available and other literature sources when not \citep[e.g.][and others]{Best2020}. For details, see the citations in Table~2 of \citet[][note that the Gaia parallaxes in that table are from DR2]{Factor2022}.

Converting angular separation to projected separation is as simple as multiplying by the distance to the target. Converting contrast into mass is not as trivial. Since brown dwarf spectral types are not a reliable mass metric, we choose to convert from absolute magnitude to bolometric luminosity using an empirical relation and finally to mass using a model dependent isochrone. 

To compute this relation, we use unresolved photometry in the NICMOS F110W and F170M filters when available (and 2MASS J and H otherwise) and convert these photometric measurements to absolute magnitudes using the distances from above. If 2MASS photometry was used, we infer the corresponding NICMOS magnitudes using an empirical relation built using synthetic photometry of the sample of brown dwarfs compiled in \citet{Filippazzo2015}. They gathered near- and mid-infrared spectra and photometry for field age and young (which we excluded) targets of spectral type M6--T9. Using geometric parallaxes, they derive bolometric luminosities for these targets and bolometric corrections for \emph{J} band photometry though, since we use HST bands, we derive our own relation based on their work.

\begin{figure*}
    \centering
    \resizebox{\textwidth}{!}{
    \includegraphics[height=3cm]{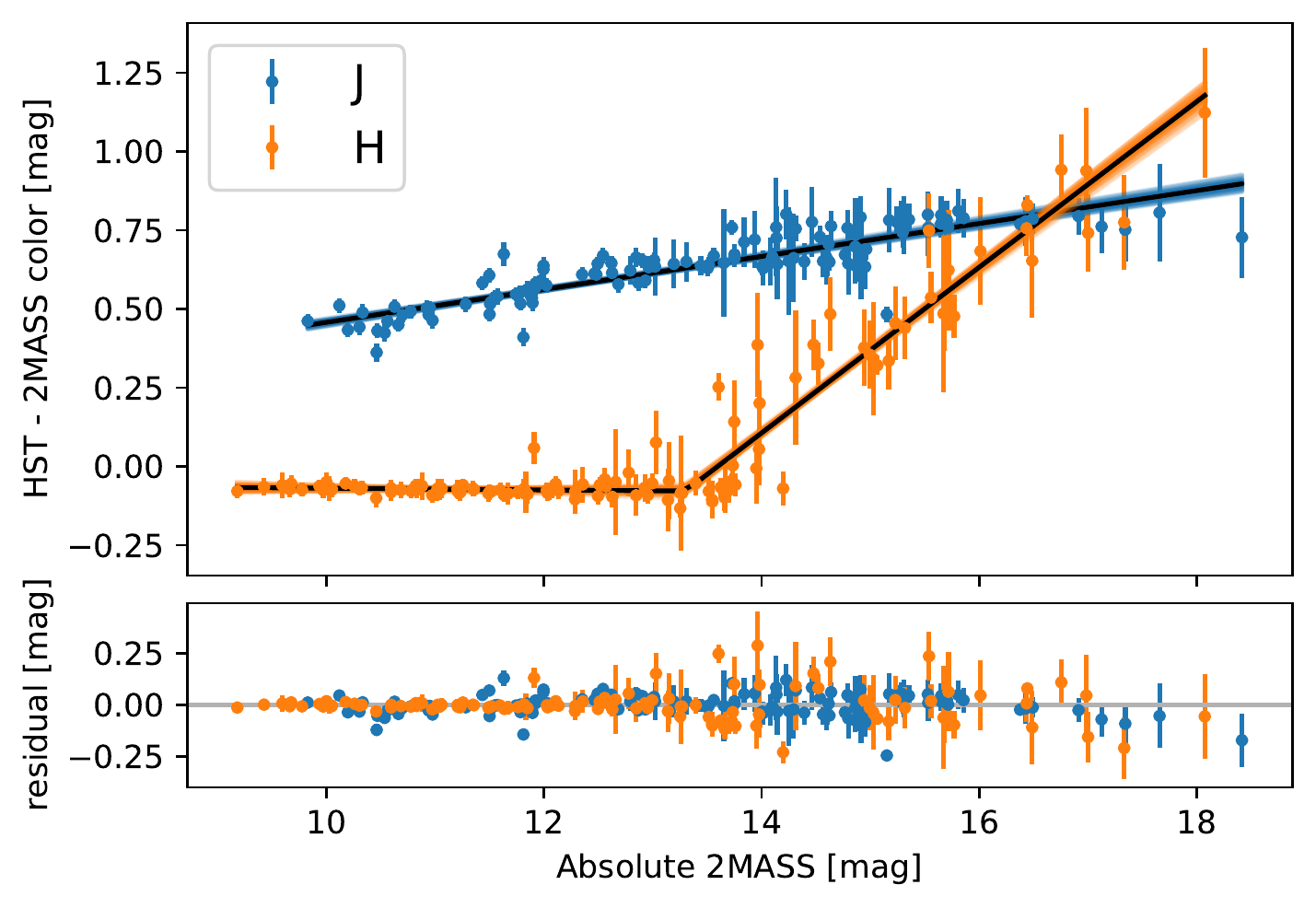}
    \includegraphics[height=3cm]{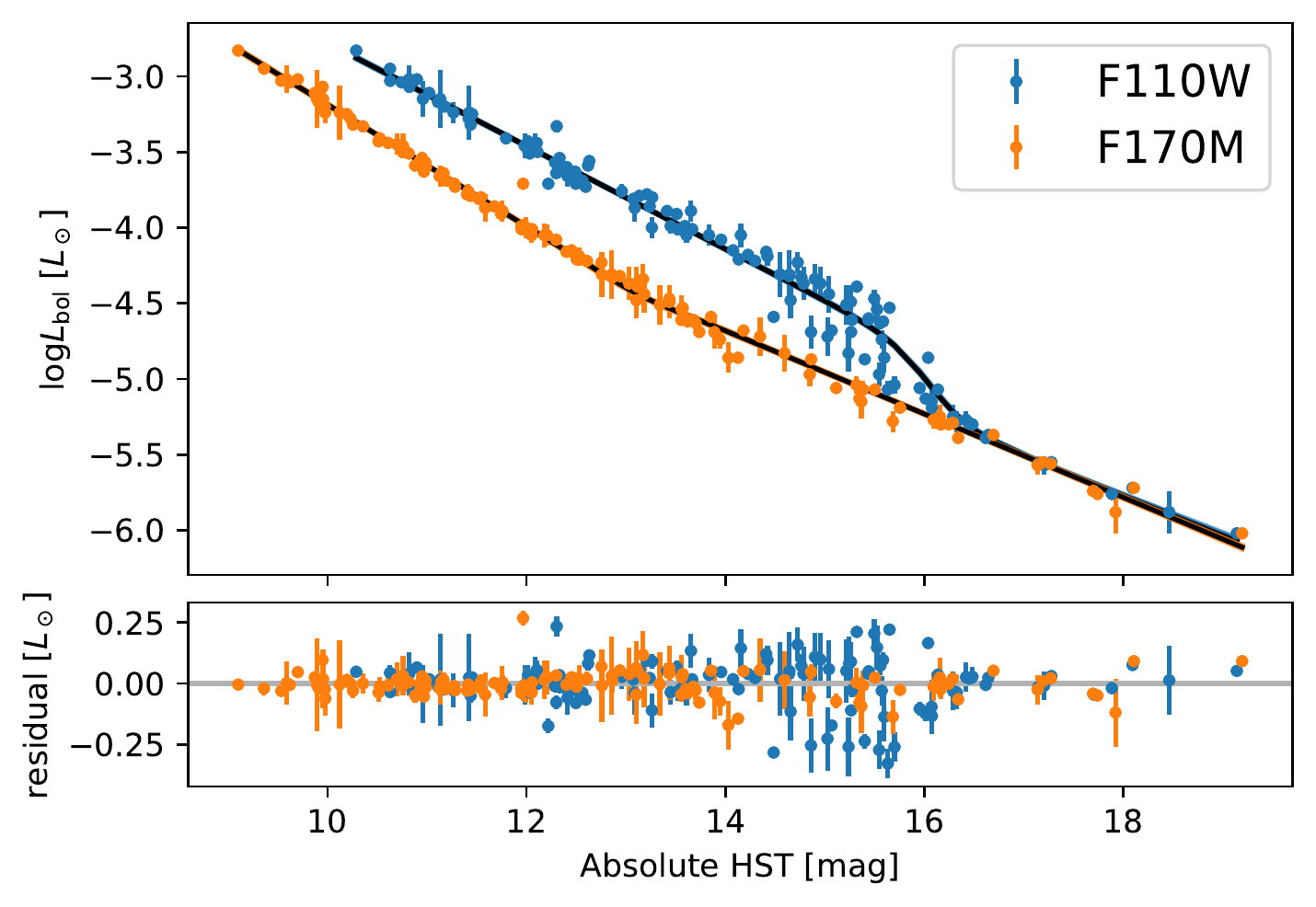}}
    \caption{Empirical relations for converting 2MASS photometry to the corresponding HST--2MASS colors (left) and NICMOS photometry to bolometric luminosity (right) built using synthetic photometry of the sample of brown dwarfs in \citet[][points and error bars]{Filippazzo2015}. Equations for the HST--2MASS colors are given in Equations~\ref{eq:phot1} and \ref{eq:phot2} and equations for HST absolute magnitude to logL are given in Equation~\ref{eq:logL} and coefficients in Table~\ref{tab:logLCoef}. Black lines are drawn using the median parameters while colored lines are 100 samples drawn from the posterior distributions. }
    \label{fig:FPhotRels}
\end{figure*}

Near infrared spectra for the empirical sample were downloaded from the BDNYC database \citep{BDNYC1,BDNYC2}. Synthetic photometry was then performed using PySynthphot \citep{pysynthphot} in the relevant HST and 2MASS filters. We then fit a broken (for \emph{H}) linear function to derive HST--2MASS color as a function of absolute 2MASS magnitude. This relation is shown in the left panel of Figure~\ref{fig:FPhotRels} and the coefficients are given by Equations~\ref{eq:phot1} and \ref{eq:phot2}.

\begin{equation}
    \mathrm{F110W}-J = 0.052\pm0.002 \times J - 0.07\pm0.02
    \label{eq:phot1}
\end{equation}
\begin{align}
    \mathrm{F170M}-H & = \left\{
    \begin{array}{ll}
    %-0.003\pm0.003 \times H - 0.04\pm0.04 \quad\quad & H<13.30\pm0.05\\
    %\phm{-} 0.263\pm0.006 \times H -3.58\pm0.09 \quad\quad & H>13.30\pm0.05
    -0.003\pm0.003 \times H - 0.04\pm0.04 \\\hfill H<13.30\pm0.05\\
    \phm{-} 0.263\pm0.006 \times H -3.58\pm0.09 \\\hfill H>13.30\pm0.05
    \end{array}\right.\label{eq:phot2}
\end{align}

We then split the unresolved (now absolute) HST band photometry into its constituent parts for a given contrast and convert these component magnitudes into bolometric luminosities using a second empirical relation also built on the \citet{Filippazzo2015} sample. Again, we fit a piecewise linear relation (this time mediated by a 5th degree sigmoid), to the bolometric luminosities as a function of absolute HST magnitude. This relation is shown in the right panel of Figure~\ref{fig:FPhotRels} and the coefficients are given by Equation~\ref{eq:logL} and Table~\ref{tab:logLCoef}. 

\begin{align}
    \log L & = (a_1 * M_\mathrm{HST} + b_1)\left(1-\frac{1}{1+e^{-5(M_\mathrm{HST}-M_0)}}\right) 
    \nonumber\\ & + (a_2* M_\mathrm{HST} - b_2)\left(\frac{1}{1+e^{-5(M_\mathrm{HST}-M_0)}}\right)
    \label{eq:logL}
\end{align}

\begin{deluxetable*}{lCCCCC}
\tabletypesize{\scriptsize}
\tablecaption{Absolute HST magnitude to LogL coefficients \label{tab:logLCoef}}
\tablehead{\colhead{Filter} & \colhead{$M_0$} & \colhead{$a_1$} & \colhead{$a_2$} & \colhead{$b_1$} & \colhead{$b_2$}}
\startdata
F110W & 16.076\pm0.016 & -0.3403\pm0.0015 & -0.263\pm0.007 &  0.623\pm0.019 & -1.04\pm0.11 \\
F170M & 13.16\pm0.13 & -0.402\pm0.003 & -0.2756\pm0.0018 &  0.83\pm0.03 & -0.82\pm0.03
\enddata
\end{deluxetable*}

From bolometric luminosity, the conversion to mass is done using an isochrone from the ATMO2020 models \citep{Phillips2020}. We derived masses using all relevant ages (0.9, 1.2, 1.5, 2.4, and 3.1~Gyr) available from the ATMO2020 models and ran our fits for each assumed field age \citep[see Section~\ref{sec:discPrev} for a more in depth discussion of the choice of field age with respect to the work of][]{Aganze2022}. We chose the ATMO2020 models due to its wide range of mass, especially at the low end, at field age ($0.075  M_\sun$ down to $0.002 M_\sun$ or $\sim79-2 M_\mathrm{Jup}$). Even though we do not detect companions at such low masses, our sensitivity does reach those masses. The median mass of our sample is $\sim65-75 M_\mathrm{Jup}$ (depending on the assumed field age) so a small amount of extrapolation is needed to extend the grid to the highest mass primaries in our sample.

\begin{figure}
    \centering
    \includegraphics[width=\columnwidth]{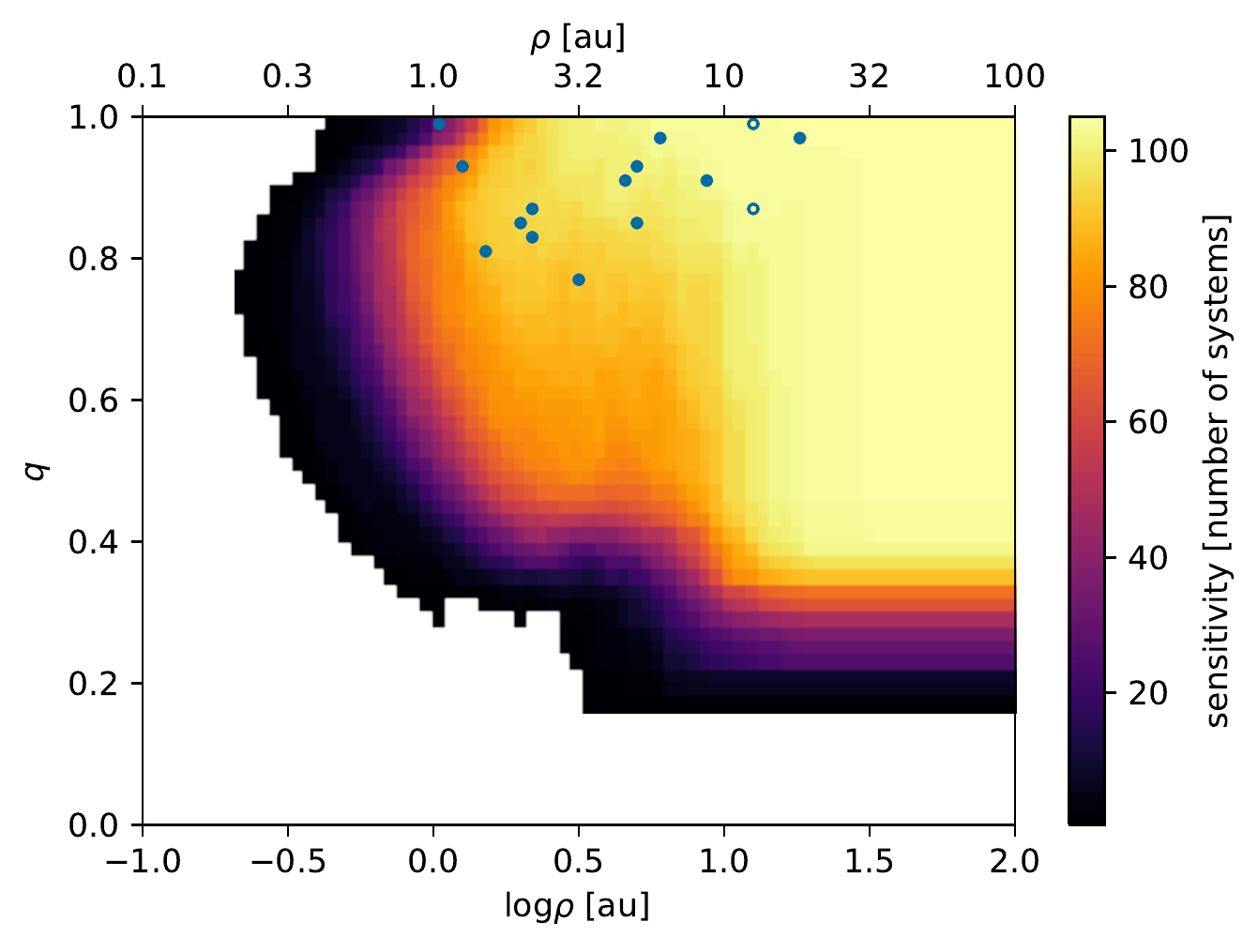}
    \caption{Stacked detection limits, in units of number of systems, as a function of projected separation, $\rho$, in au and mass-ratio, $q$, using a field age of 1.9~Gyr \citep[similar to Figure~15 of][though now in physical units rather than observational units]{Factor2022}. Filled circles show detected companions and open circles indicate the two wide separation companions not detected in \citet{Factor2022} but included in our sample using astrometry and photometry from \citet{Pope2013} (see Section~\ref{sec:nicmos}). We set a conservative detection limit at separations $>0\farcs5$ (where our KPI pipeline is not sensitive) of $\Delta m = 6.5$ from \citet{Reid2006}. Corresponding plots for other field ages are available online at \dataset[DOI:10.5281/zenodo.7370349]{https://doi.org/10.5281/zenodo.7370349} \citet{figSets2}.}
    \label{fig:totSens}
\end{figure}

Fitted separations and contrasts \citep[both from][]{Factor2022}, and observed and intermediary derived properties of detected binaries are given in Table \ref{tab:binPropInt}. Projected separations and masses (and mass ratios) are given in Table~\ref{tab:binPropFin} and plotted on top of the stacked sensitivity of the \citet{Factor2022} analysis in Figure~\ref{fig:totSens} for a given assumed field age (and \dataset[DOI:10.5281/zenodo.7370349]{https://doi.org/10.5281/zenodo.7370349} \citep{figSets2} for other assumed ages). As noted in \citet{Reid2006}, due to the similar slopes of the isochrones in $L_\mathrm{bol}$-Mass space, a higher $q$ is inferred when using an older age.

\begin{deluxetable*}{lCCCCCCCC}
\tabletypesize{\tiny}
%\tabletypesize{\scriptsize}
    \tablewidth{0pt}
    \tablecaption{Fit and Intermediary Binary Properties\label{tab:binPropInt}}
    %\tablehead{\colhead{Source} & \colhead{distance} & \colhead{$\rho$} & \nocolhead{$\rho$} & \colhead{$m_\mathrm{F110W}$} & \colhead{$m_\mathrm{F170M}$} & \colhead{$c_\mathrm{F110W}$} & \colhead{$c_\mathrm{F170M}$} & \colhead{$\log L_\mathrm{A}$} & \colhead{$\log L_\mathrm{B}$} & \nocolhead{$q_{0.9}$} & \nocolhead{$M_\mathrm{A,0.9}$} & \nocolhead{$M_\mathrm{B,0.9}$} & \nocolhead{$q_{1.2}$} & \nocolhead{$M_\mathrm{A,1.2}$} & \nocolhead{$M_\mathrm{B,1.2}$} & \nocolhead{$q_{1.5}$} & \nocolhead{$M_\mathrm{A,1.5}$} & \nocolhead{$M_\mathrm{B,1.5}$} & \nocolhead{$q_{1.9}$} & \nocolhead{$M_\mathrm{A,1.9}$} & \nocolhead{$M_\mathrm{B,1.9}$} & \nocolhead{$q_{2.4}$} & \nocolhead{$M_\mathrm{A,2.4}$} & \nocolhead{$M_\mathrm{B,2.4}$} & \nocolhead{$q_{3.1}$} & \nocolhead{$M_\mathrm{A,3.1}$} & \nocolhead{$M_\mathrm{B,3.1}$}\\
    %\colhead{} & \colhead{[pc]} & \colhead{[mas]} & \nocolhead{[au]} & \colhead{[mag]} & \colhead{[mag]} & \colhead{} & \colhead{} & \colhead{[$L_\odot$]} & \colhead{[$L_\odot$]} & \nocolhead{} & \nocolhead{[$M_\mathrm{Jup}]$} & \nocolhead{[$M_\mathrm{Jup}$]} & \nocolhead{} & \nocolhead{[$M_\mathrm{Jup}]$} & \nocolhead{[$M_\mathrm{Jup}$]} & \nocolhead{} & \nocolhead{[$M_\mathrm{Jup}]$} & \nocolhead{[$M_\mathrm{Jup}$]} & \nocolhead{} & \nocolhead{[$M_\mathrm{Jup}]$} & \nocolhead{[$M_\mathrm{Jup}$]} & \nocolhead{} & \nocolhead{[$M_\mathrm{Jup}]$} & \nocolhead{[$M_\mathrm{Jup}$]} & \nocolhead{} & \nocolhead{[$M_\mathrm{Jup}]$} & \nocolhead{[$M_\mathrm{Jup}$]}}
    \tablehead{\colhead{Source} & \colhead{distance} & \colhead{$\rho$} & \colhead{$m_\mathrm{F110W}$} & \colhead{$m_\mathrm{F170M}$} & \colhead{$c_\mathrm{F110W}$} & \colhead{$c_\mathrm{F170M}$} & \colhead{$\log L_\mathrm{A}$} & \colhead{$\log L_\mathrm{B}$} \\
    \colhead{} & \colhead{[pc]} & \colhead{[mas]} & \colhead{[mag]} & \colhead{[mag]} & \colhead{} & \colhead{} & \colhead{[$L_\odot$]} & \colhead{[$L_\odot$]}}
\startdata
%       source &      dist &        sep &        m110 &        m170 &             c0 &                       c1 &       logLA &       logLB \\
 2M 0004-4044 &  12.13 \pm 0.06\tablenotemark{1} &     83.2 \pm 0.5 &                     13.80 \pm 0.05 &                     12.10 \pm 0.05 &      1.126 \pm 0.018 &  \mathrm{nan} \pm \mathrm{nan} &  -4.161 \pm 0.018 &  -4.205 \pm 0.018 \\
 2M 0025+4759 &    54.0 \pm 0.4\tablenotemark{2} &    334.5 \pm 0.9 &                     15.56 \pm 0.05 &                     13.55 \pm 0.05 &      1.353 \pm 0.033 &                1.251 \pm 0.023 &  -3.533 \pm 0.014 &  -3.637 \pm 0.014 \\
 2M 0147-4954 &    34.7 \pm 0.4\tablenotemark{1} &    138.8 \pm 0.4 &                     13.27 \pm 0.05 &                     13.08 \pm 0.05 &      2.345 \pm 0.030 &                2.022 \pm 0.016 &  -3.251 \pm 0.015 &  -3.562 \pm 0.015 \\
 2M 0423-0414 &  14.28 \pm 0.19\tablenotemark{1} &  159.11 \pm 0.24 &                     15.28 \pm 0.05 &                     13.62 \pm 0.05 &      1.655 \pm 0.011 &                2.124 \pm 0.011 &  -4.466 \pm 0.017 &  -4.683 \pm 0.016 \\
 2M 0429-3123 &  17.00 \pm 0.05\tablenotemark{1} &        534 \pm 4 &                     11.23 \pm 0.05 &                     10.17 \pm 0.05 &        3.50 \pm 0.23 &                  2.77 \pm 0.13 &  -2.919 \pm 0.014 &  -3.375 \pm 0.018 \\
 2M 0700+3157 &  11.23 \pm 0.05\tablenotemark{1} &    179.4 \pm 0.7 &                     13.17 \pm 0.05 &                     11.27 \pm 0.05 &        4.43 \pm 0.12 &                3.809 \pm 0.033 &  -3.779 \pm 0.013 &  -4.343 \pm 0.015 \\
 2M 0915+0422 &    17.7 \pm 0.3\tablenotemark{1} &  738.60 \pm 0.15 &                     15.30 \pm 0.05 &                     13.57 \pm 0.05 &  1.11686 \pm 0.00020 &            1.08643 \pm 0.00020 &  -4.380 \pm 0.017 &  -4.418 \pm 0.017 \\
 2M 0926+5847 &    23.0 \pm 0.5\tablenotemark{3} &   67.22 \pm 0.14 &                     16.57 \pm 0.05 &                     15.64 \pm 0.05 &      1.522 \pm 0.018 &                  2.70 \pm 0.28 &  -4.685 \pm 0.018 &  -4.971 \pm 0.026 \\
 2M 1021-0304 &    29.7 \pm 1.1\tablenotemark{3} &    166.4 \pm 0.5 &                     17.09 \pm 0.05 &                     15.83 \pm 0.05 &      1.104 \pm 0.015 &                2.516 \pm 0.015 &  -4.639 \pm 0.021 &  -4.856 \pm 0.022 \\
 2M 1553+1532 &  13.32 \pm 0.16\tablenotemark{4} &    345.7 \pm 0.7 &                     16.55 \pm 0.04 &                     16.43 \pm 0.04 &      1.363 \pm 0.034 &                1.408 \pm 0.027 &  -5.341 \pm 0.011 &  -5.451 \pm 0.010 \\
 2M 1707-0558 &  11.95 \pm 0.03\tablenotemark{1} &   1009.5 \pm 1.0 &                     12.59 \pm 0.05 &                     11.25 \pm 0.05 &        5.15 \pm 0.15 &                  3.34 \pm 0.20 &  -3.629 \pm 0.014 &  -4.206 \pm 0.017 \\
 2M 2152+0937 &        24 \pm 4\tablenotemark{5} &    254.2 \pm 0.6 &                     16.02 \pm 0.05 &                     14.02 \pm 0.05 &      1.157 \pm 0.019 &                1.126 \pm 0.012 &    -4.36 \pm 0.10 &    -4.41 \pm 0.10 \\
 2M 2252-1730 &  16.53 \pm 0.16\tablenotemark{1} &    126.7 \pm 0.7 &                     15.10 \pm 0.05 &                     13.46 \pm 0.05 &      2.568 \pm 0.028 &                  3.21 \pm 0.04 &  -4.246 \pm 0.016 &  -4.637 \pm 0.013 \\
 2M 2255-5713 &  17.07 \pm 0.16\tablenotemark{1} &    178.3 \pm 0.8 &  14.694 \pm 0.030\tablenotemark{a} &  13.115 \pm 0.032\tablenotemark{a} &        5.16 \pm 0.14 &                  4.53 \pm 0.06 &  -4.057 \pm 0.010 &  -4.646 \pm 0.010 \\
 2M 2351-2537 &  20.34 \pm 0.19\tablenotemark{1} &   62.65 \pm 0.34 &                     12.90 \pm 0.05 &                     11.84 \pm 0.05 &        2.58 \pm 0.09 &                    2.9 \pm 0.9 &  -3.376 \pm 0.017 &  -3.720 \pm 0.020
\enddata
\tablerefs{1: \citet{BailerJones2021}, 2: \citet{Gaia2018}, 3: \citet{Dupuy2017}, 4: \citet{Dupuy2012}, 5: \citet{Best2020}}
\tablenotemark{a}{HST band photometry derived from 2MASS photometry. }
\end{deluxetable*}

\begin{deluxetable*}{lChhhCCChhhhhhhhhCCC}
\tabletypesize{\tiny}
%\tabletypesize{\scriptsize}
    \tablewidth{0pt}
    \tablecaption{Final Binary Properties \label{tab:binPropFin}}
    \tablehead{\colhead{Source} & \colhead{$\rho$} & \nocolhead{$q_{0.9}$} & \nocolhead{$M_\mathrm{A,0.9}$} & \nocolhead{$M_\mathrm{B,0.9}$} & \colhead{$q_{1.2}$} & \colhead{$M_\mathrm{A,1.2}$} & \colhead{$M_\mathrm{B,1.2}$} & \nocolhead{$q_{1.5}$} & \nocolhead{$M_\mathrm{A,1.5}$} & \nocolhead{$M_\mathrm{B,1.5}$} & \nocolhead{$q_{1.9}$} & \nocolhead{$M_\mathrm{A,1.9}$} & \nocolhead{$M_\mathrm{B,1.9}$} & \nocolhead{$q_{2.4}$} & \nocolhead{$M_\mathrm{A,2.4}$} & \nocolhead{$M_\mathrm{B,2.4}$} & \colhead{$q_{3.1}$} & \colhead{$M_\mathrm{A,3.1}$} & \colhead{$M_\mathrm{B,3.1}$}\\
    \colhead{} & \colhead{[au]} & \nocolhead{} & \nocolhead{[$M_\mathrm{Jup}]$} & \nocolhead{[$M_\mathrm{Jup}$]} & \colhead{} & \colhead{[$M_\mathrm{Jup}]$} & \colhead{[$M_\mathrm{Jup}$]} & \nocolhead{} & \nocolhead{[$M_\mathrm{Jup}]$} & \nocolhead{[$M_\mathrm{Jup}$]} & \nocolhead{} & \nocolhead{[$M_\mathrm{Jup}]$} & \nocolhead{[$M_\mathrm{Jup}$]} & \nocolhead{} & \nocolhead{[$M_\mathrm{Jup}]$} & \nocolhead{[$M_\mathrm{Jup}$]} & \colhead{} & \colhead{[$M_\mathrm{Jup}]$} & \colhead{[$M_\mathrm{Jup}$]}}
\startdata
 2M 0004-4044 &  1.013 \pm 0.008 &  0.970 \pm 0.017 &    61.4 \pm 0.7 &    59.5 \pm 0.8 &  0.977 \pm 0.013 &    66.2 \pm 0.6 &    64.7 \pm 0.6 &  0.982 \pm 0.010 &    70.2 \pm 0.5 &    68.9 \pm 0.5 &  0.987 \pm 0.008 &    72.9 \pm 0.4 &    71.9 \pm 0.4 &    0.990 \pm 0.006 &  74.71 \pm 0.29 &  73.97 \pm 0.31 &    0.992 \pm 0.004 &  75.87 \pm 0.24 &  75.28 \pm 0.25 \\
 2M 0025+4759 &   18.06 \pm 0.14 &  0.959 \pm 0.008 &    82.8 \pm 0.5 &    79.4 \pm 0.5 &  0.966 \pm 0.006 &    84.1 \pm 0.4 &    81.3 \pm 0.4 &  0.973 \pm 0.005 &  84.41 \pm 0.31 &  82.12 \pm 0.31 &  0.978 \pm 0.004 &  84.33 \pm 0.25 &  82.47 \pm 0.25 &  0.9812 \pm 0.0035 &  84.25 \pm 0.21 &  82.66 \pm 0.22 &  0.9839 \pm 0.0030 &  83.94 \pm 0.18 &  82.59 \pm 0.18 \\
 2M 0147-4954 &    4.98 \pm 0.07 &  0.891 \pm 0.007 &    91.9 \pm 0.5 &    81.8 \pm 0.5 &  0.907 \pm 0.006 &    91.9 \pm 0.4 &    83.3 \pm 0.4 &  0.925 \pm 0.005 &  90.56 \pm 0.33 &  83.77 \pm 0.33 &  0.938 \pm 0.004 &  89.30 \pm 0.27 &  83.80 \pm 0.27 &    0.947 \pm 0.004 &  88.49 \pm 0.23 &  83.80 \pm 0.23 &  0.9542 \pm 0.0031 &  87.57 \pm 0.20 &  83.56 \pm 0.20 \\
 2M 0423-0414 &    2.34 \pm 0.05 &  0.848 \pm 0.018 &    49.1 \pm 0.7 &    41.6 \pm 0.6 &  0.852 \pm 0.016 &    54.6 \pm 0.7 &    46.5 \pm 0.6 &  0.858 \pm 0.014 &    60.2 \pm 0.7 &    51.7 \pm 0.6 &  0.877 \pm 0.011 &    64.9 \pm 0.6 &    56.9 \pm 0.6 &    0.905 \pm 0.009 &    68.6 \pm 0.4 &    62.1 \pm 0.5 &    0.930 \pm 0.007 &  71.18 \pm 0.32 &    66.2 \pm 0.4 \\
 2M 0429-3123 &    8.99 \pm 0.07 &  0.856 \pm 0.007 &   102.6 \pm 0.5 &    87.9 \pm 0.6 &  0.876 \pm 0.006 &   101.0 \pm 0.4 &    88.5 \pm 0.5 &  0.898 \pm 0.005 &  97.83 \pm 0.31 &    87.9 \pm 0.4 &  0.915 \pm 0.004 &  95.19 \pm 0.25 &  87.12 \pm 0.32 &    0.926 \pm 0.004 &  93.51 \pm 0.21 &  86.62 \pm 0.27 &  0.9359 \pm 0.0031 &  91.86 \pm 0.18 &  85.98 \pm 0.23 \\
 2M 0700+3157 &  2.032 \pm 0.011 &  0.724 \pm 0.009 &    74.7 \pm 0.5 &    54.1 \pm 0.6 &  0.771 \pm 0.009 &    77.4 \pm 0.4 &    59.6 \pm 0.6 &  0.818 \pm 0.007 &  79.01 \pm 0.29 &    64.6 \pm 0.5 &  0.858 \pm 0.005 &  79.96 \pm 0.24 &    68.6 \pm 0.4 &    0.887 \pm 0.004 &  80.52 \pm 0.20 &  71.42 \pm 0.29 &  0.9076 \pm 0.0034 &  80.76 \pm 0.17 &  73.29 \pm 0.23 \\
 2M 0915+0422 &   13.48 \pm 0.27 &  0.973 \pm 0.018 &    52.6 \pm 0.6 &    51.2 \pm 0.7 &  0.974 \pm 0.016 &    58.1 \pm 0.7 &    56.6 \pm 0.7 &  0.980 \pm 0.013 &    63.3 \pm 0.6 &    62.1 \pm 0.6 &  0.983 \pm 0.010 &    67.6 \pm 0.5 &    66.5 \pm 0.5 &    0.988 \pm 0.008 &    70.6 \pm 0.4 &    69.8 \pm 0.4 &    0.991 \pm 0.006 &  72.69 \pm 0.28 &  72.06 \pm 0.30 \\
 2M 0926+5847 &    1.54 \pm 0.04 &  0.812 \pm 0.022 &    41.6 \pm 0.7 &    33.8 \pm 0.7 &  0.813 \pm 0.020 &    46.4 \pm 0.7 &    37.7 \pm 0.7 &  0.814 \pm 0.018 &    51.6 \pm 0.6 &    42.0 \pm 0.8 &  0.819 \pm 0.017 &    56.9 \pm 0.6 &    46.6 \pm 0.8 &    0.831 \pm 0.016 &    62.1 \pm 0.5 &    51.6 \pm 0.9 &    0.857 \pm 0.015 &    66.2 \pm 0.4 &    56.7 \pm 0.9 \\
 2M 1021-0304 &    4.94 \pm 0.18 &  0.842 \pm 0.017 &    43.2 \pm 0.7 &    36.4 \pm 0.4 &  0.844 \pm 0.016 &    48.0 \pm 0.7 &    40.5 \pm 0.5 &  0.848 \pm 0.017 &    53.1 \pm 0.7 &    45.1 \pm 0.6 &  0.858 \pm 0.017 &    58.5 \pm 0.7 &    50.1 \pm 0.8 &    0.878 \pm 0.016 &    63.4 \pm 0.6 &    55.7 \pm 0.9 &    0.907 \pm 0.014 &    67.3 \pm 0.5 &    61.0 \pm 0.8 \\
 2M 1553+1532 &    4.60 \pm 0.06 &  0.913 \pm 0.011 &  24.69 \pm 0.23 &  22.55 \pm 0.18 &  0.914 \pm 0.011 &  27.72 \pm 0.26 &  25.35 \pm 0.20 &  0.915 \pm 0.011 &  31.13 \pm 0.29 &  28.47 \pm 0.22 &  0.915 \pm 0.011 &  34.96 \pm 0.32 &  31.99 \pm 0.24 &    0.917 \pm 0.010 &  39.18 \pm 0.33 &  35.91 \pm 0.28 &    0.918 \pm 0.010 &  43.83 \pm 0.34 &  40.24 \pm 0.31 \\
 2M 1707-0558 &   11.88 \pm 0.06 &  0.747 \pm 0.010 &    79.7 \pm 0.4 &    59.5 \pm 0.7 &  0.793 \pm 0.008 &    81.5 \pm 0.4 &    64.6 \pm 0.6 &  0.837 \pm 0.007 &  82.30 \pm 0.30 &    68.9 \pm 0.5 &  0.871 \pm 0.005 &  82.62 \pm 0.24 &    71.9 \pm 0.4 &    0.893 \pm 0.004 &  82.79 \pm 0.21 &  73.96 \pm 0.29 &  0.9101 \pm 0.0034 &  82.70 \pm 0.18 &  75.26 \pm 0.23 \\
 %2M 2028+0052 &    1.11 \pm 0.10 &    0.90 \pm 0.26 &        71 \pm 9 &       64 \pm 17 &    0.92 \pm 0.22 &        75 \pm 8 &       69 \pm 15 &    0.94 \pm 0.19 &        77 \pm 6 &       72 \pm 13 &    0.95 \pm 0.16 &        78 \pm 5 &       75 \pm 11 &      0.96 \pm 0.13 &        79 \pm 4 &       76 \pm 10 &      0.97 \pm 0.11 &    79.4 \pm 3.4 &        77 \pm 8 \\
 2M 2152+0937 &      6.2 \pm 1.1 &    0.96 \pm 0.10 &        53 \pm 4 &        52 \pm 4 &    0.97 \pm 0.09 &        59 \pm 4 &        57 \pm 4 &    0.97 \pm 0.08 &    64.1 \pm 3.4 &        62 \pm 4 &    0.98 \pm 0.06 &    68.2 \pm 2.8 &    66.8 \pm 3.0 &      0.98 \pm 0.04 &    71.1 \pm 2.1 &    70.0 \pm 2.3 &    0.989 \pm 0.032 &    73.1 \pm 1.6 &    72.2 \pm 1.7 \\
 2M 2252-1730 &  2.143 \pm 0.032 &  0.749 \pm 0.011 &    57.8 \pm 0.6 &    43.3 \pm 0.4 &  0.760 \pm 0.009 &    63.2 \pm 0.5 &    48.1 \pm 0.4 &  0.786 \pm 0.009 &    67.7 \pm 0.5 &    53.2 \pm 0.5 &  0.824 \pm 0.008 &    71.0 \pm 0.4 &    58.5 \pm 0.5 &    0.866 \pm 0.006 &  73.25 \pm 0.28 &    63.5 \pm 0.4 &    0.901 \pm 0.005 &  74.71 \pm 0.22 &  67.32 \pm 0.32 \\
 2M 2255-5713 &  3.030 \pm 0.033 &  0.661 \pm 0.006 &  65.16 \pm 0.35 &  43.05 \pm 0.34 &  0.687 \pm 0.005 &  69.56 \pm 0.30 &  47.77 \pm 0.32 &  0.727 \pm 0.005 &  72.78 \pm 0.23 &  52.89 \pm 0.35 &  0.777 \pm 0.005 &  74.95 \pm 0.19 &    58.2 \pm 0.4 &    0.828 \pm 0.004 &  76.32 \pm 0.15 &  63.19 \pm 0.30 &  0.8694 \pm 0.0035 &  77.19 \pm 0.12 &  67.11 \pm 0.24 \\
 2M 2351-2537 &  1.275 \pm 0.014 &  0.873 \pm 0.009 &    87.8 \pm 0.6 &    76.7 \pm 0.7 &  0.893 \pm 0.008 &    88.4 \pm 0.5 &    79.0 \pm 0.6 &  0.914 \pm 0.006 &    87.8 \pm 0.4 &    80.3 \pm 0.4 &  0.930 \pm 0.005 &  87.09 \pm 0.30 &    81.0 \pm 0.4 &    0.940 \pm 0.005 &  86.61 \pm 0.26 &  81.41 \pm 0.31 &    0.948 \pm 0.004 &  85.96 \pm 0.22 &  81.52 \pm 0.26
\enddata
\tablecomments{1.2 and 3.1 refer to the assumed age of the field population when converting magnitude or bolometric luminosity to mass. Masses and mass ratios calculated using other ages considered in this work are given in the online table.}
\end{deluxetable*}

\subsection{Binary Population}\label{sec:binPop}
We apply a similar Bayesian modeling approach to \citet{Kraus2011} and \citet{Kraus2012} \citep[adapted from][]{Allen2007} to infer the underlying binary population from our sample of observations. This technique is particularly suited to the relatively small number of detections at close separations since there is no need to throw out data to create a volume limited sample or correct for regions of incompleteness. However, we must still account for Malmquist bias in our magnitude limited sample.

The binary population is characterized by a companion frequency $F$, a power-law mass ratio distribution with exponent $\gamma$, and a log-normal projected separation distribution with mean $\overline{\log(\rho)}$ and standard deviation $\sigma_{\log(\rho)}$. As in \citet{Kraus2011}, we choose to model the separation distribution in terms of the observed projected separation ($\rho$) rather than the underlying semimajor axis ($a$)\footnote{This avoids making an assumption about the eccentricity distribution which may change with future observations. Converting $\rho$ to $a$ can be done using a simple conversion factor calculated using montecarlo simulations of projected orbits. For very low-mass binaries $a/\rho=1.16^{+0.81}_{-0.31}$, for no discovery bias, or as low as $a/\rho=0.85^{+0.11}_{-0.14}$, for a survey with an inner working angle comparable to $a$ \citep{Dupuy2011}.}. The binary population model is then used in our likelihood function by comparing it to the observations over a grid $(\Delta\log(\rho),\Delta q)$ in parameter space. The expected companion frequency in a bin is given by the probability $R$ according to: 
\begin{align}
    R&(\log(\rho),q|F,\gamma,\overline{\log(\rho)},\sigma_{\log(\rho)})\Delta\log\rho\Delta q = \frac{\gamma + 1}{\sqrt{2\pi}\sigma_{\log(\rho)}} 
    \nonumber\\ & \times Fq^\gamma\exp\left(-\frac{(\log(\rho)-\overline{\log(\rho)})^2}{2\sigma^2_{\log(\rho)}}\right)\Delta\log\rho\Delta q \label{eq:compfreq}
\end{align}

Before this companion frequency is used in the likelihood function, we must first include Malmquist bias \citep{Malmquist1922} in order to compare it to our magnitude limited sample. In previous studies this bias correction was applied to the detection limits by artificially increasing the sensitivity to binaries since they should be over represented in the underlying sample. However, this method does not account for unresolved binaries which are treated as single sources. Instead, we build Malmquist bias into our population model by increasing the number of binaries, as a function of contrast, that we should expect to observe (though not necessarily resolve). Since our model is a function of mass-ratio $q=M_\mathrm{B}/M_\mathrm{A}$ and Malmquist bias is a function of total brightness or contrast $C=F_\mathrm{A}/F_\mathrm{B}$, we must first convert $q$ to $C$. We chose the contrast in F110W ($C_\mathrm{F110W}$) since it closely resembles the 2MASS-$J$ filter used to select the sample. Since we already convert $C$ to $q$, as described above, the inverse is numerically trivial. We calculate the Malmquist correction using the same method as \citet{Burgasser2003b} and \citet{Allen2007},
\begin{equation}
    V_\mathrm{max}/V_\mathrm{flux}=\left(1+\frac{1}{C_\mathrm{F110W}}\right)^{3/2},\label{eq:malm}
\end{equation}
for each source then average to apply a single correction to the model. We ran the fit with and without the Malmquist correction in order to both determine the underlying and unbiased population and compare our results to previous analysis of ``observed" populations.

The Bayesian likelihood in a given bin $(\log(\rho),q)$ with $N_\mathrm{det}$ companions detected and $N_\mathrm{sen}$ targets which were sensitive to such a companion is then the Binomial likelihood:
%\begin{equation}
%    P(N_\mathrm{det},N_\mathrm{sen}|F,\gamma,\overline{\log(\rho)},\sigma_{\log(\rho)})\propto R^{N_\mathrm{det}} \times (1-R)^{(N_\mathrm{sen}-N_\mathrm{det})} \label{eq:like}.
%\end{equation}
\begin{align}
    P&(N_\mathrm{det},N_\mathrm{sen}|F,\gamma,\overline{\log(\rho)},\sigma_{\log(\rho)})
    \nonumber\\&\propto R^{N_\mathrm{det}} \times (1-R)^{(N_\mathrm{sen}-N_\mathrm{det})} \label{eq:like}.
\end{align}
The grid is derived from the overall survey detections and sensitivity and is shown in Figure \ref{fig:totSens} along with the binary systems. The full grid spans $\log(\rho)=-4$ to 3 in 175 bins and $q=0$ to 1 in 50 bins. Since our calculated limits only extend to 0.5 arcsec we adopt a 100\% detection threshold of $\Delta m=6.5$ magnitudes for all separations $>0.5$ arcsec. This is a conservative estimate of the sensitivity of the \citet{Reid2006} analysis. The $\log(\rho)$ axis of our grid is much larger than the region occupied by our detections (our detections span a range of $\sim75$ bins) to give our detection limits (where we are sensitive but did not detect any companions) and informed prior (see Section~\ref{sec:demoI}) leverage over the model parameters. 

This likelihood function is then passed to a fitting routine, in our case \texttt{emcee} \citep{Foreman-Mackey2013} which implemented the affine invariant sampler described in \citet{Goodman2010}. We ran the fit using 64 ``walkers" for 10000 steps (discarding the first 1000 for ``burn-in"), exceeding 50 autocorrelation times for all four parameters while only running for a few minutes.

We first ran the fit with uninformed (wide and flat) priors on all four parameters ($F=0$ to 2.5, $\gamma=-1$ to 20, $\overline{\log(\rho)}=-4$ to 3, and $\sigma_{\log(\rho)}=0$ to 4). This produced an interesting binary distribution, though one that is inconsistent with previous RV studies. Thus, we ran our fits a second time with an informed prior (though technically implemented as a penalized likelihood), restricting the tight ($<1~\mathrm{au}$) binary fraction to $2.5^{+8.6}_{-1.6}\%$ \citep{Blake2010}. This was implemented by summing up $R$ (Equation~\ref{eq:compfreq}) for all projected separation bins $<1$~au (resulting in the tight companion fraction) and penalizing the likelihood based on the difference between this sum and the measurement of \citet{Blake2010}. This resulted in a much more tightly constrained posterior on the population distribution parameters. \citet{Blake2010} estimated a 94\% completeness for targets with $0.01 < a < 1$~au while our completeness is $\sim50\%$ at 1~au and drops to $\sim0\%$ by 0.3~au. While the measurement by \citet{Blake2010} assumed a much flatter mass-ratio distribution \citep[$\gamma=1.8$ from][]{Allen2007}, they claim their sensitivity (and thus their measurement of binary fraction) is relatively insensitive to the distribution. We apply this prior to the underlying companion distribution, marginalizing over $q$, so their choice of $\gamma$ will have little effect on our results.

\section{Results: Binary Demographics}\label{sec:res}

Previous studies of brown dwarf demographics using a similar model (log-normal separation and power-law mass-ratio distributions) have found a population heavily skewed toward equal mass ($\gamma\sim2-5$) with a semimajor axis distribution centered around $\sim6$ au and a total companion frequency of $\sim20\%$ \citep{Reid2006,Burgasser2007,Allen2007}. \citet{Fontanive2018} studied later spectral type objects (T5--Y0) and found a roughly similar semimajor axis distribution with a lower companion frequency of $8\pm6\%$ and a slightly stronger mass-ratio power-law index of $\gamma\sim6$. Radial velocity studies searching for tight companions (separation $<1~au$) have found a much lower companion frequency of $2.5^{+8.6}_{-1.6}\%$ \citep{Blake2010} and $2\pm2\%$ for late type objects \citep{Fontanive2018}. Initially we ran our demographic fits using a wide and uninformed prior. Since our detections run up against our inner working angle, these fits yielded a companion distribution with an extremely high binary fraction at tight separations, inconsistent with previous studies. We therefore incorporated an informed prior, restricting the amount of tight binaries, and found a population much more consistent with previous studies. Since the age and mass of a brown dwarf are degenerate (for a given absolute magnitude) we also ran all of our fits with a set of assumed field ages. This only affected the mass-ratio power-law index and had no effect on the total companion frequency or semimajor axis parameters. We ran our fits with and without the Malmquist bias correction in order to study both the underlying unbiased population and the ``observed" binary population, to compare with previous studies. Incorporating Malmquist bias reduced both the companion frequency, $F$, and mass-ratio power-law index, $\gamma$, while having no effect on the separation distribution parameters, $\overline{\log(\rho)}$ and $\sigma_{\log(\rho)}$. $F$ decreases because binaries are over-represented in the observed sample, while $\gamma$ decreases since equal-brightness (i.e. high $q$) systems are more over-represented then those with fainter (i.e. lower $q$) companions.

\subsection{Demographics from an Uninformed Prior}\label{sec:demoU}
\begin{figure*}
    \centering
    \includegraphics[width=\textwidth]{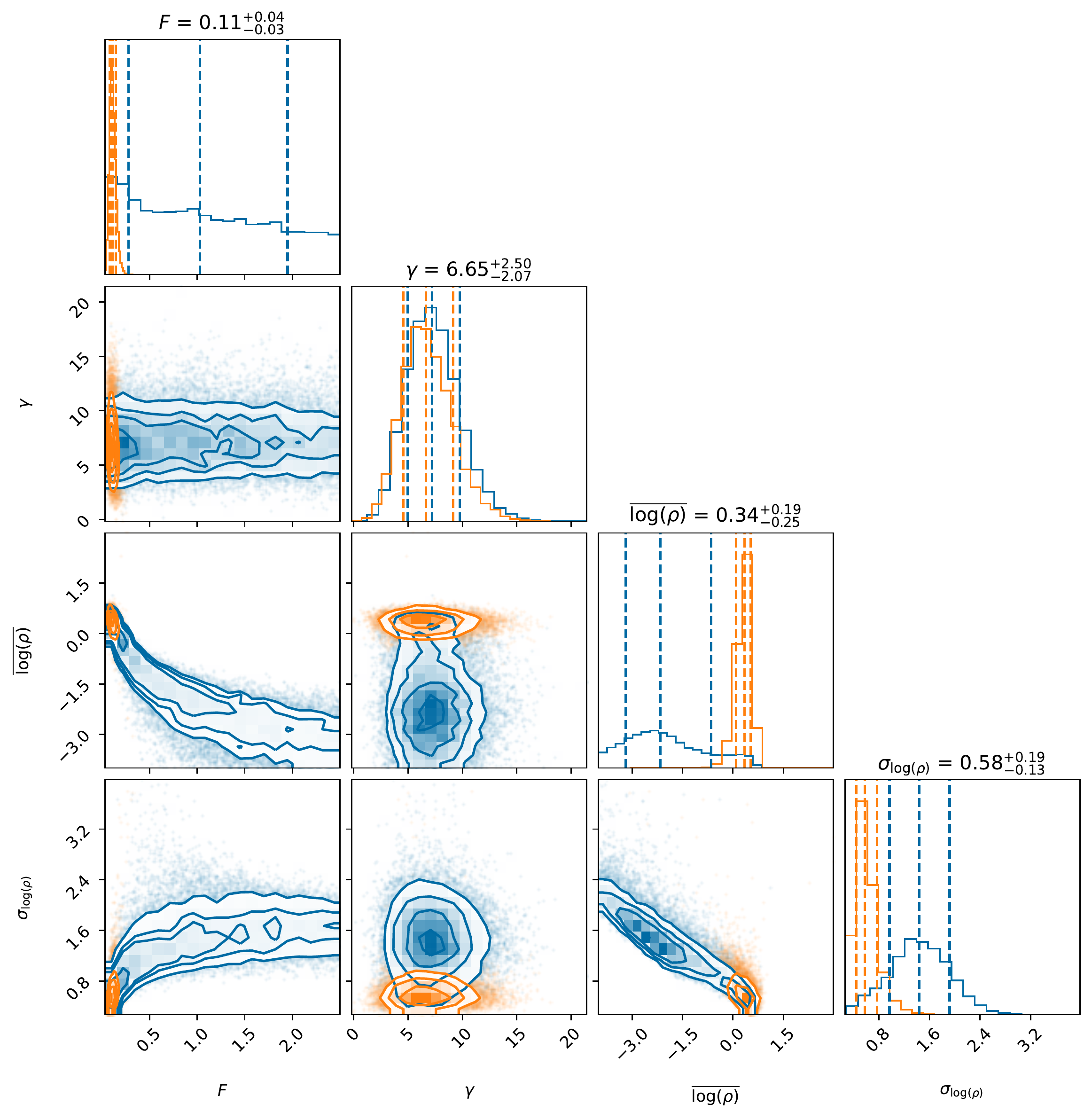}
    \caption{Corner plot showing the 1- and 2D posteriors of our demographic fit using a field age of 1.9~Gyr. Blue contours show the results using a uniform uninformed prior while orange contours show the results restricting the tight ($<1$~au) binary fraction to $2.5^{+8.6}_{-1.6}\%$ \citet{Blake2010}. Dashed lines indicate the median and $\pm1\sigma$ (16th, 50th, and 84th percentile) values (given in Table~\ref{tab:pop}). The parameter values listed above each plot correspond to the fits using the informed prior. Corresponding plots for other field ages are available online at \dataset[DOI:10.5281/zenodo.7370349]{https://doi.org/10.5281/zenodo.7370349} \citet{figSets2}. Data behind the figure is available for the MCMC chains shown in this figure and those for other ages at \dataset[DOI:10.5281/zenodo.7065651]{https://doi.org/10.5281/zenodo.7065651} \cite{mcmcChains}.}
    \label{fig:1p9GyrCorner}
\end{figure*}

Posterior distributions using an uninformed prior are shown as a corner plot in blue in Figure~\ref{fig:1p9GyrCorner} for an assumed field age of 1.9~Gyr (and \dataset[DOI:10.5281/zenodo.7370349]{https://doi.org/10.5281/zenodo.7370349} \citep{figSets2} for other ages). Median and central 68\% credible intervals for the model hyper-parameters are given in Table~\ref{tab:pop}. Since age only affects the derived mass and has no bearing on separation, the only parameter affected by the assumed field age is $\gamma$ and it is discussed below. 

\begin{deluxetable}{cCCCC}
    \tablecaption{Binary Population Parameters\label{tab:pop}}
    \tablehead{\colhead{Isochrone} & \colhead{$\gamma$} & \colhead{$F$} & \colhead{$\overline{\log(\rho)}$} & \colhead{$\sigma_{\log(\rho)}$} }
\startdata
\multicolumn{5}{l}{Informed Prior, Underlying Population}\\
   3.1 Gyr & 11^{+4}_{-3} & 0.11^{+0.04}_{-0.03} & 0.34^{+0.18}_{-0.25} & 0.58^{+0.20}_{-0.13} \\
   2.4 Gyr & 9\pm3 & 0.11^{+0.04}_{-0.03} & 0.34^{+0.19}_{-0.25} & 0.58^{+0.19}_{-0.13} \\
   1.9 Gyr & 7^{+3}_{-2} & 0.11^{+0.04}_{-0.03} & 0.34^{+0.19}_{-0.25} & 0.58^{+0.19}_{-0.13} \\
   1.5 Gyr & 5.5^{+2.1}_{-1.8} & 0.11^{+0.04}_{-0.03} & 0.34^{+0.19}_{-0.25} & 0.58^{+0.20}_{-0.13} \\
   1.2 Gyr & 4.6^{+1.8}_{-1.6} & 0.11^{+0.04}_{-0.03} & 0.34^{+0.19}_{-0.24} & 0.57^{+0.20}_{-0.13} \\
   0.9 Gyr & 4.0^{+1.7}_{-1.5} & 0.11^{+0.04}_{-0.03} & 0.34^{+0.19}_{-0.25} & 0.58^{+0.19}_{-0.13} \\
\multicolumn{5}{l}{Informed Prior, Observed Population\tablenotemark{a}}\\
   3.1 Gyr & 14\pm4 & 0.22^{+0.07}_{-0.06} & 0.35^{+0.18}_{-0.25} & 0.58^{+0.20}_{-0.13} \\
   2.4 Gyr & 11\pm3 & 0.22^{+0.07}_{-0.06} & 0.33^{+0.19}_{-0.26} & 0.58^{+0.20}_{-0.13} \\
   1.9 Gyr & 9^{+3}_{-2} & 0.22^{+0.07}_{-0.05} & 0.35^{+0.18}_{-0.25} & 0.57^{+0.20}_{-0.13} \\
   1.5 Gyr & 7\pm2 & 0.22^{+0.07}_{-0.05} & 0.35^{+0.18}_{-0.25} & 0.57^{+0.20}_{-0.13} \\
   1.2 Gyr & 6.2^{+2.0}_{-1.7} & 0.22^{+0.07}_{-0.06} & 0.35^{+0.19}_{-0.25} & 0.57^{+0.20}_{-0.13} \\
   0.9 Gyr & 5.5^{+1.8}_{-1.6} & 0.22^{+0.07}_{-0.06} & 0.34^{+0.19}_{-0.25} & 0.57^{+0.20}_{-0.13} \\
\multicolumn{5}{l}{Uninformed Prior, Underlying Population}\\
   3.1 Gyr & 12^{+4}_{-3} & 1.0^{+0.9}_{-0.8}\tablenotemark{b} & -2.2^{+1.5}_{-1.0}\tablenotemark{c} & 1.5\pm0.5 \\
   2.4 Gyr & 10\pm3 & 1.1^{+0.9}_{-0.8}\tablenotemark{b} & -2.2^{+1.5}_{-1.0}\tablenotemark{c} & 1.5\pm0.5 \\
   1.9 Gyr & 7^{+3}_{-2} & 1.0^{+0.9}_{-0.8}\tablenotemark{b} & -2.2^{+1.5}_{-1.0}\tablenotemark{c} & 1.4\pm0.5 \\
   1.5 Gyr & 5.9^{+2.2}_{-1.9} & 1.0^{+0.9}_{-0.8}\tablenotemark{b} & -2.1^{+1.5}_{-1.1}\tablenotemark{c} & 1.4\pm0.5 \\
   1.2 Gyr & 4.9^{1.9}_{-1.6} & 1.0^{+0.9}_{-0.8}\tablenotemark{b} & -2.2^{+1.5}_{-1.0}\tablenotemark{c} & 1.4\pm0.5 \\
   0.9 Gyr & 4.3^{+1.8}_{-1.5} & 1.0^{+0.9}_{-0.8}\tablenotemark{b} & -2.1^{+1.5}_{-1.0}\tablenotemark{c} & 1.4\pm0.5 \\
\enddata
\tablecomments{Values presented here are median and central 68\% credible intervals.}
\tablenotetext{a}{The ``observed" population parameters were fit without correcting for Malmquist bias.}
\tablenotetext{b}{The median and central 68\% confidence intervals are not good metrics as the posterior on F with the uninformed prior is relatively flat over the span of the allowed parameters with a weak peak near the same value as with the informed prior (see Figure~\ref{fig:1p9GyrCorner}).}
\tablenotetext{c}{The posterior on $\overline{\log(\rho)}$ with the uninformed prior has a weak secondary peak near the same value as with the informed prior (see Figure~\ref{fig:1p9GyrCorner}).}
\end{deluxetable}

The uninformed prior resulted in a posterior distribution on companion frequency which is peaked at small $F$ with a long tail extending to extremely high companion fraction. Companion frequency is also degenerate with a broad (large $\sigma_{\log(\rho)}$) separation distribution centered at $\rho\sim0.006$~au. While there is a weak peak at $F\sim0.1$ and $\overline{\log(\rho)}\sim0.3$ (likely corresponding to the true underlying distribution, see Section~\ref{sec:demoI}) it is overwhelmed by a distribution containing a large number of undetectable companions inside the inner working angle of our survey. In order for a survey to rule out unresolved companions the inner working angle must fall well inside the peak of the separation distribution. If this is the case, a decrease in detections can then be attributed to the population rather than a cutoff in sensitivity. Since our survey includes detections that run up against the inner working angle ($\sim1$~au), we cannot rule out a significant population of unresolved companions. Thus, a companion distribution centered at $\overline{\log(\rho)}=-2.2$ or $\rho=0.006$~au (well within our inner working angle) must also be wide ($\sigma_{\log(\rho)}=1.4$) and requires an extremely high companion fraction to reproduce the observations since our observed companions are located in the upper wing of the distribution. 

We note a strong degeneracy between $F$ and both separation parameters: $\overline{\log(\rho)}$ and $\sigma_{\log(\rho)}$. This was also seen in previous studies \citep[e.g.][]{Allen2007,Kraus2011}. This degeneracy can be explained since a companion distribution centered at a tighter separation ($\overline{\log(\rho)}$) needs both a higher companion fraction ($F$) and a wider distribution ($\sigma_{\log(\rho)}$) to reproduce the observed companions which are now further out in the wing of the distribution. This also explains the strong degeneracy between $\overline{\log(\rho)}$ and $\sigma_{\log(\rho)}$. \citet{Allen2007} also noted a degeneracy between $F$ and $\overline{\log(\rho)}$ attributed to their inability to rule out a population of tight, unresolved, binaries. With a sufficiently large sample the curvature in the wing of the distribution can be used to break this degeneracy (assuming the underlying distribution is truly log-normal) but we are nowhere close to having enough detections for this to be possible. 

\subsection{Demographics from an Informed Prior}\label{sec:demoI}
\begin{figure*}
    \centering
    \includegraphics[width=\textwidth]{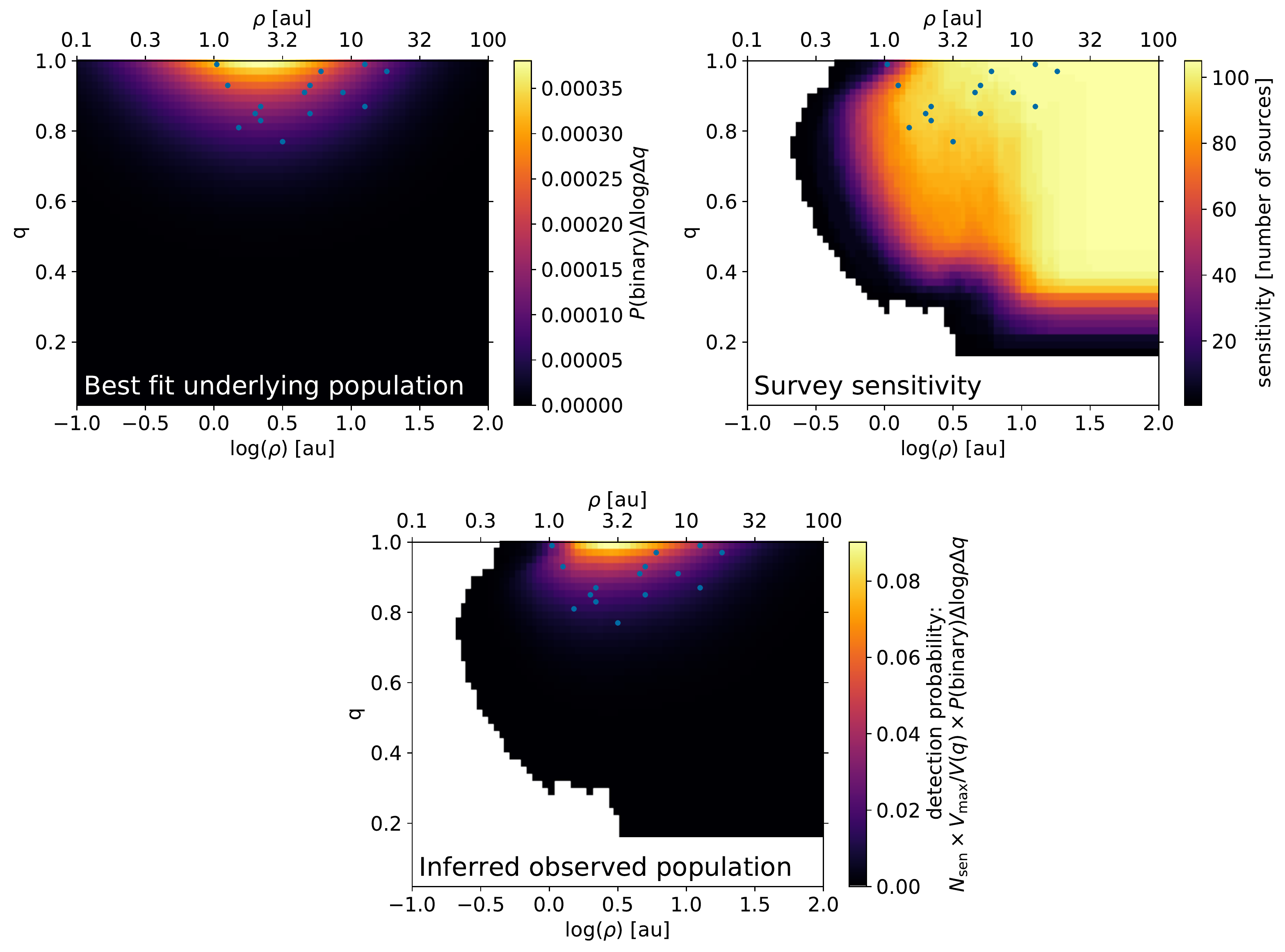}
    \caption{The progression from underlying population and survey sensitivity to observed population. Blue points indicate detected companions. \emph{Top left:} Underlying companion population produced from the median values of our informed prior fit. \emph{Top right:} Survey sensitivity in units of number of targets. \emph{Bottom center:} Observed population (i.e. the inferred probability that we should detect a companion in a given bin), calculated by correcting the underlying population for Malmquist bias (as a function of mass-ratio) and applying our survey sensitivity. Mass ratios were calculated assuming a field age of 1.9~Gyr. Similar figures for other assumed field ages are available online at \dataset[DOI:10.5281/zenodo.7370349]{https://doi.org/10.5281/zenodo.7370349} \citep{figSets2}.}
    \label{fig:2dpop}
\end{figure*}

While the demographics derived using the uninformed prior are consistent with our observations they are not consistent with previous studies of brown dwarf binarity at separations inside our inner working angle. \citet{Blake2010} used six years of near-infrared RV measurements of 50 late-M and L dwarfs from Keck/NIRSPEC to infer the tight ($<1~\mathrm{au}$) binary fraction to be $2.5^{+8.6}_{-1.6}\%$. This binary fraction is significantly lower than the $140\pm100\%$ (essentially every system is a binary or triple) our posterior parameters would predict if marginalized over the same range in separation and inflated to simulate an observed sample (accounting for Malmquist bias). Thus, we ran a second set of fits restricting the tight binary fraction to be consistent with that of \citet{Blake2010} (as discussed at the end of Section~\ref{sec:binPop}). The distribution produced while considering the \citet{Blake2010} information is much more consistent with previous studies. The tight ($<1$~au) binary fraction of our posterior binary BD population (again, corrected for Malmquist bias) is $6^{+6}_{-4}\%$, slightly higher than though well within the $1\sigma$ error bars of \citet{Blake2010}. To the best of our knowledge, this is the first time an RV prior has been incorporated in a direct imaging BD demographics survey to account for unresolved companions. The use of this prior is further discussed in Section~\ref{sec:disc}. The posteriors produced by this fit are shown in orange in Figure~\ref{fig:1p9GyrCorner} and median values and central 68\% credible intervals for the model hyper-parameters are given in Table~\ref{tab:pop}. The full 2D binary population is shown in the left panel of Figure~\ref{fig:2dpop} beside the observed population (correcting for Malmquist bias and sensitivity) and our raw sensitivity along with our detections. 

\begin{figure}
    \centering
    \includegraphics[width=\columnwidth]{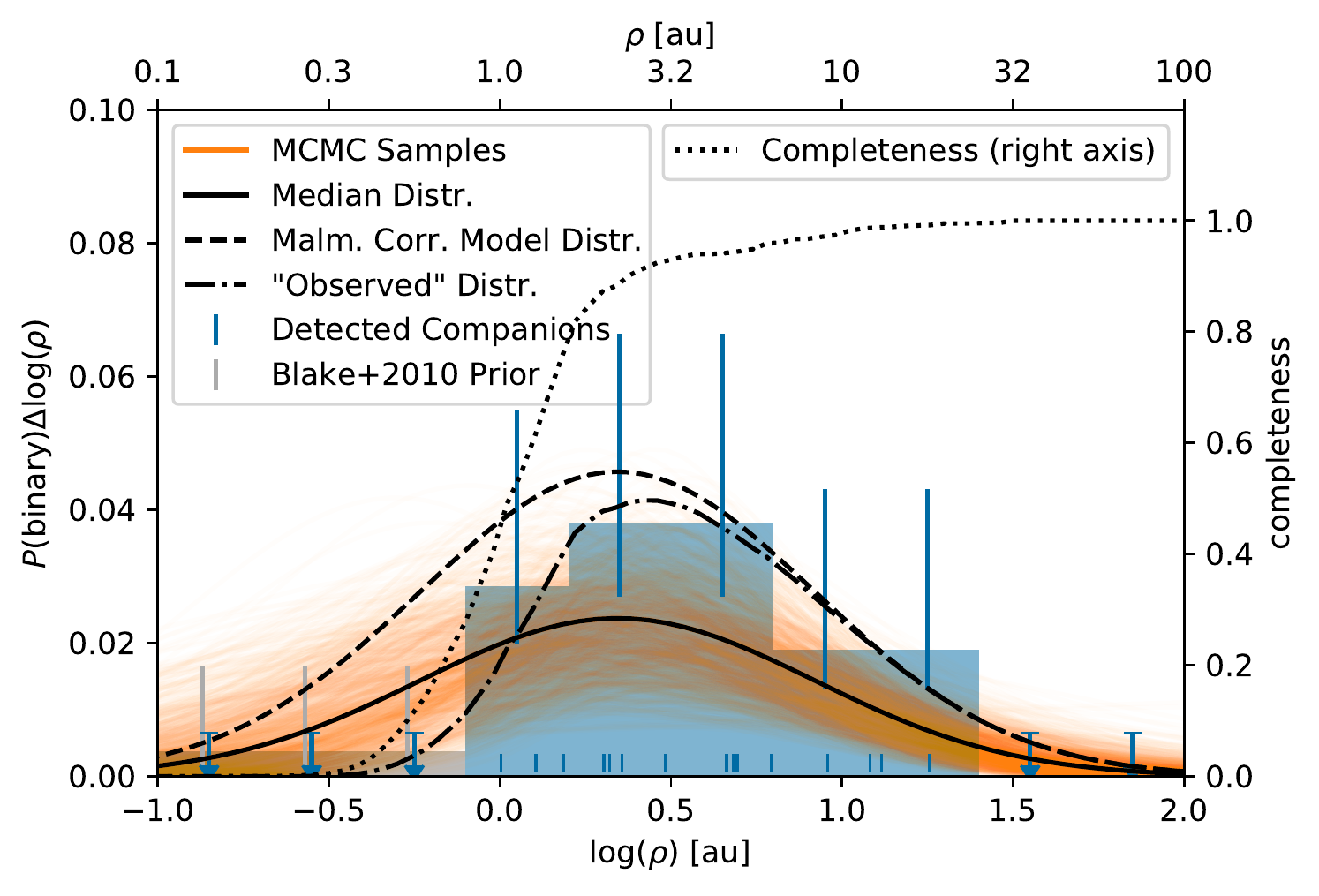}
    \caption{Binary population as a function of projected separation. The blue histogram shows the detected companion separations and error bars calculated using \citet{Burgasser2003b} with dashes below showing the un-binned values. Grey bars and error bars correspond to the \citet{Blake2010} measurement of tight separation binaries ($2.5^{+8.6}_{-1.6}\%$ spread out evenly over $\log(\rho)=-2$ to 0). Orange curves are 1000 companion probability densities drawn from the posterior distributions using the informed prior while the solid black curve is drawn using the median parameters ($F=0.11^{+0.04}_{-0.03}$, $\overline{\log(\rho)}=0.34^{+0.19}_{-0.25}$, $\sigma_{\log(\rho)}=0.58^{+0.20}_{-0.13}$, i.e the top left panel of Figure~\ref{fig:2dpop} marginalized over $q$). The black dashed line shows the Malmquist corrected distribution (see Section~\ref{sec:binPop}) and the dash-dotted line shows the ``observed" distribution (bottom center pannel of Figure~\ref{fig:2dpop}), calculated by multiplying the Malmquist corrected binary-population distribution by our sensitivity (shown in Figure \ref{fig:totSens} or the top right panel of Figure~\ref{fig:2dpop}). The black dotted line is the completeness fraction (right vertical axis), the ratio between the black dashed and dash-dotted lines or equivalently the marginalized and normalized sensitivity. }
    \label{fig:binslogp}
\end{figure}

The observed and inferred binary probability density as a function of projected separation (marginalized over mass ratio $q$) is shown in Figure~\ref{fig:binslogp}. Histogram error-bars are drawn according to \citet{Burgasser2003b} and agree well with the ``observed" distribution drawn by correcting the companion distribution (calculated using the median-fit hyper parameters) for Malmquist bias and applying the survey sensitivity. The increase in observed binaries above the underlying distribution is due to Malmquist bias while the drop in observed binaries at close separations ($<1$~au) is due to a steep drop in sensitivity. Also shown is the information from \citet{Blake2010} used to constrain the unresolved population ($2.5^{+8.6}_{-1.6}\%$ spread out evenly over $\log(\rho)=-2$ to 0.  

\subsection{Mass-ratio Demographics}\label{sec:demoM}
Since the mass of a brown dwarf depends on the assumed age, we re-ran the demographic fit with multiple assumed field ages. The information incorporated in the informed prior only affects tight-binary fraction and has no information about mass ratio, thus the posteriors on $\gamma$ are essentially the same when using the informed or uninformed prior. Similarly, as age only affects the derived mass (or mass ratio $q$), other model parameters do not significantly change as a function of age. 

\begin{figure}
    \centering
    \includegraphics[width=\columnwidth]{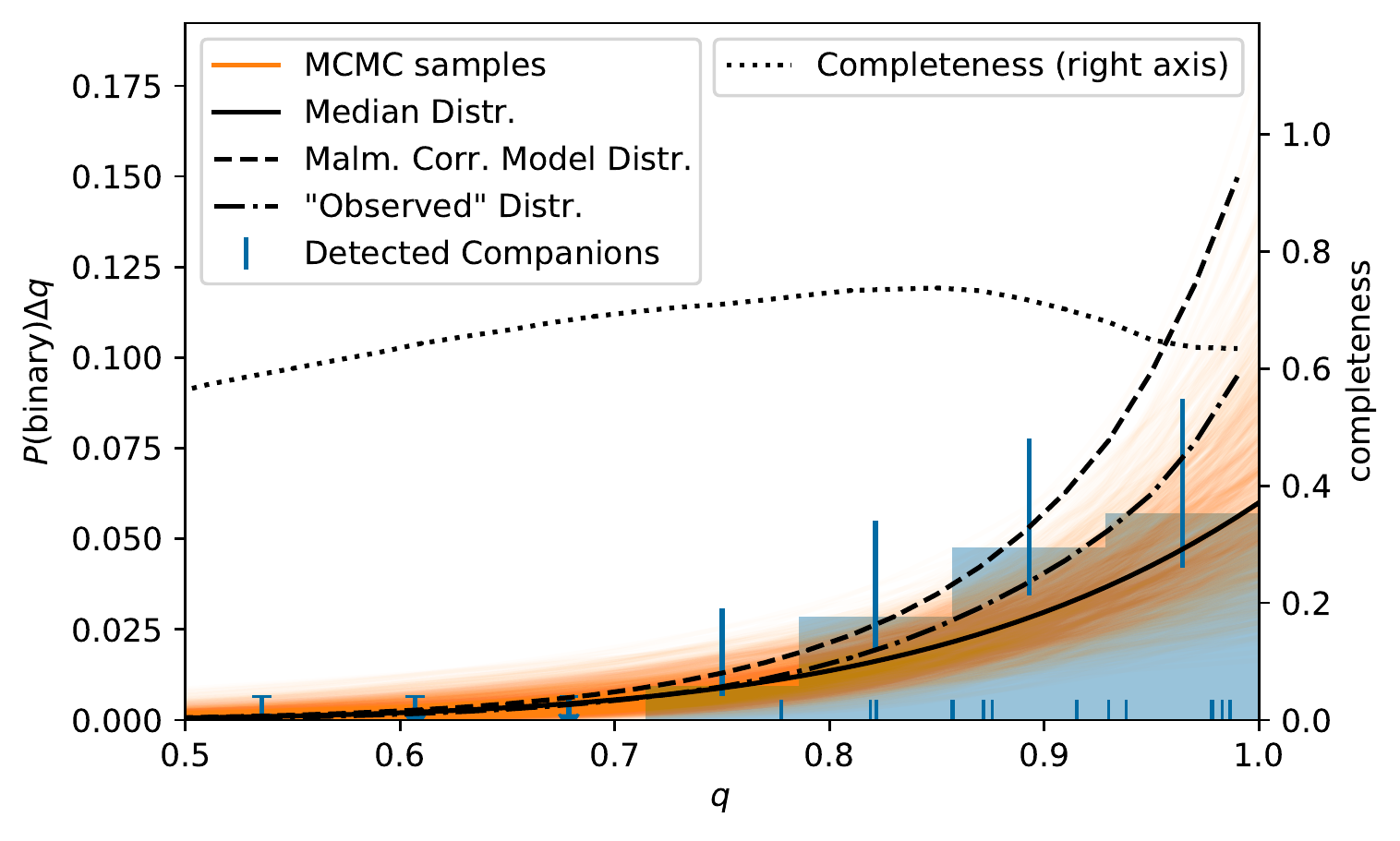}
    \caption{Similar to Figure~\ref{fig:binslogp} but as a function of mass ratio. Mass ratios are derived using a 1.9~Gyr assumed field age and a median power-law index of $\gamma=7^{+3}_{-2}$ and overall companion frequency of $F=0.11^{+0.04}_{-0.03}$. Similar figures for other assumed field ages are available online at \dataset[DOI:10.5281/zenodo.7370349]{https://doi.org/10.5281/zenodo.7370349} \citet{figSets2}.}
    \label{fig:binsq}
\end{figure}

Figure~\ref{fig:binsq} shows the binary fraction as a function of mass ratio for an assumed field age of 1.9~Gyr (other ages are available online at \dataset[DOI:10.5281/zenodo.7370349]{https://doi.org/10.5281/zenodo.7370349} \citep{figSets2}). Once again, the ``observed" distribution (median posterior distribution corrected for Malmquist bias times sensitivity) matches well with the histogram of detected companions. As opposed to the separation distribution (shown in Figure~\ref{fig:binslogp}), our sensitivity is relatively constant well beyond the region of parameter space inhabited by our detections so no additional information is needed to prevent the fit from producing distributions with a large population of faint undetectable companions.

\begin{figure}
    \centering
    \includegraphics[width=\columnwidth]{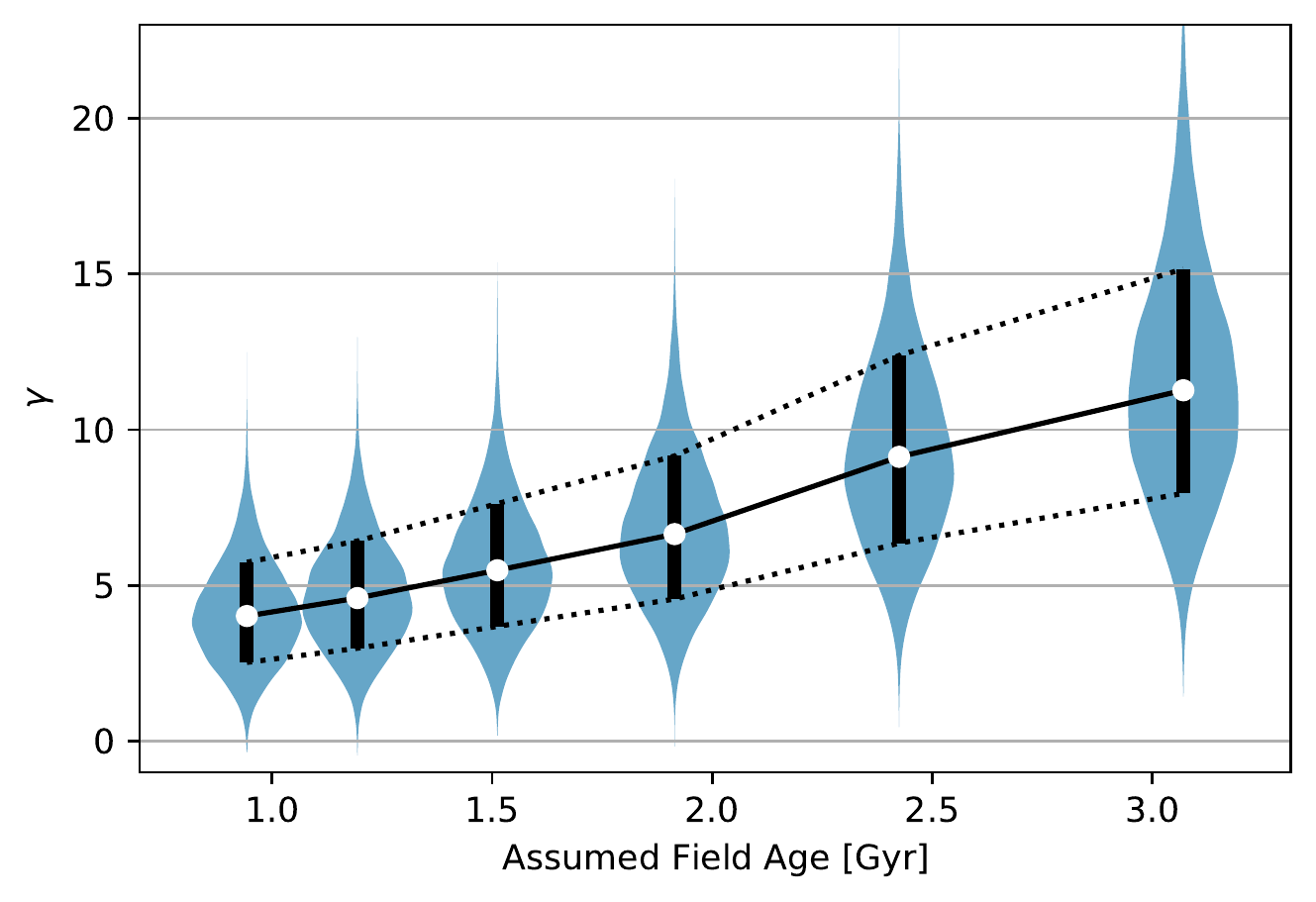}
    \caption{Posteriors of the mass-function power-law index $\gamma$ as a function of assumed field age. White points and black vertical lines correspond to median and $\pm1\sigma$ values.}
    \label{fig:gammaAge}
\end{figure}

As stated above, we ran our demographic fit for a variety of field ages ranging from 0.9--3.1~Gyr. Since age only affects the mass (or mass ratio) of a companion and not the separation, the only demographic parameter that changes with age is $\gamma$ (and it is not affected by the choice of prior on the amount of tight separation binaries). Figure~\ref{fig:gammaAge} shows the posterior distributions on $\gamma$ as a function of the assumed field age. Median values and central 68\% credible intervals are given in Table~\ref{tab:pop} and range from $\gamma=4.0^{+1.7}_{-1.5}$ to $\gamma=11^{+4}_{-3}$ assuming an age of 0.9 or 3.1~Gyr, respectively.

\subsection{``Observed" Population Without Accounting for Malmquist Bias}

In order to compare our results to literature studies which report parameters of the \emph{observed} population we run our fits a final time without accounting for Malmquist bias. The results of these fits are presented in Table~\ref{tab:pop}. The separation distribution hyper parameters are essentially identical since Malmquist bias corrects for the number of binaries (resolved or unresolved) as a function of contrast (or mass ratio $q$) which are included in a magnitude limited sample (due to their increased brightness). Accordingly $F$ is much larger when Malmquist bias is not accounted for as companions are naturally over-represented in the observed sample. The mass-ratio power-law index, $\gamma$, is roughly $1\sigma$ larger when Malmquist bias is not accounted for as equal brightness (high $q$) companions are more over-represented than faint (low $q$) companions.

\section{Discussion}\label{sec:disc}

\subsection{Demographics in the Context of Previous Surveys}\label{sec:discPrev}

Our BD binary demographic parameters are largely consistent with the previous literature values. We compare our values with three studies of early type BDs and one of late type BDs: the original study on a subset of our data set \citep{Reid2006}, a meta-analysis of BD binary studies \citep{Allen2007}, a review of BD binarity \citep{Burgasser2007}, and a study of late T and Y dwarfs of even lower mass than included in our sample \citep{Fontanive2018}. Using our sample of 15 detected binaries in 105 targets \citep[$\epsilon_b = 14^{+4}_{-3}\%$, with error bars from binomial statistics following][]{Burgasser2003b} and fitting for the full population as described in Section~\ref{sec:res}, we infer an underlying companion frequency of $F=0.11^{+0.04}_{-0.03}$. ``Observing" this population by multiplying the full 2D companion distribution by our detection limits and applying a Malmquist bias correction gives a predicted number of observed companions of $15^{+4}_{-3}$ and an observed companion frequency of $14^{+4}_{-3}\%$, both consistent with our observed value. Fitting the ``observed" population gives a frequency of $F=0.22^{+0.07}_{-0.06}$ (from the fits ran without accounting for Malmquist bias), consistent with the values of the three studies of L dwarfs ($0.24^{+0.06}_{-0.02}$, $0.20\pm0.04$, and $0.22^{+0.08}_{-0.04}$, respectively) to well within $1\sigma$, while our unbiased (Malmquist bias corrected) underlying companion frequency is 2--3$\sigma$ lower than these values. Since \citet{Reid2006} does not account for Malmquist bias this is not surprising. On the other hand \citet{Allen2007} and \citet{Burgasser2007}, who uses the same method, both account for Malmquist bias in their sensitivity window function but arrive at a similar companion frequency, $F\sim0.22$, to that of \citet{Reid2006}. \citet{Fontanive2018} reports a binary fraction of $F=0.08\pm0.06$, continuing the trend of decreasing binary fraction with decreasing primary mass. 

Our mean separation ($\overline{\log(\rho)}=0.34^{+0.19}_{-0.25}$ or $\overline{\rho}=2.2^{+1.2}_{-1.0}$~au) is tighter than the values from the above L dwarf studies ($\overline{\log(\rho)}=0.8^{+0.06}_{-0.12}$, $0.86^{+0.06}_{-0.12}$, and $0.86^{+0.06}_{-0.18}$, respectively, or $\overline{\rho}\sim6-7$~au) by $\sim2\sigma$, while our separation standard-deviation ($\sigma_{\log(\rho)}=0.58^{+0.2}_{-0.13}$) is larger than the above studies ($\sigma_{\log(\rho)}=0.28\pm0.4$, $0.28\pm0.04$, and $0.24^{+0.08}_{-0.06}$, respectively) by $0.7-2.2\sigma$. While our mean separation is consistent (at $\sim0.3\sigma$) with the value reported in \citet[][$\overline{\log(\rho)}=2.9^{+0.8}_{-1.4}$]{Fontanive2018}, their $\sigma_{\log(\rho)}=0.21^{+0.14}_{-0.08}$ is again smaller by $1.9\sigma$. Since our KPI technique has a much smaller inner working angle and our analysis includes one new tight companion (2M 2351-2537, discovered by \citet{Pope2013} and confirmed by \citet{Factor2022}), it is not surprising that our separation distribution has moved in. Our separation distribution must then also be wider to stay consistent with the widely separated companions. The information in our ``informed prior" \citep[from the RV survey by][]{Blake2010} keeps the mean from moving even closer in. Also worth noting is the fact that we used distances measured from geometric parallaxes \citep{BailerJones2021,Gaia2021} to convert observed separations (in arcsec) to projected separations (in au) rather than spectroscopic parallaxes used the studies discussed above \citep[except for][who used a combination of the two methods]{Fontanive2018}. Geometric parallaxes are more accurate and precise then spectroscopic parallaxes. The Malmquist correction had no effect on these parameters.

Most previous BD demographic studies used restricted priors to avoid the tight separation/high companion frequency solution that we found using an uninformed prior in Section~\ref{sec:demoU}. Of the four previously discussed studies, \citet{Fontanive2018} is the only one which explicitly lists the bounds of their prior. They use a flat prior that is relatively narrow around the expected value, effectively eliminating the tight median separation solutions. While using a flat prior with a limited range is effective at rejecting unreasonable solutions, the bounds can be arbitrary. If a wide flat prior produces unphysical results (such as a population with $a\sim R_\mathrm{BD}$) a more robust strategy is to establish where the data or likelihood function is lacking leverage and use a physically- or observationally-motivated informed prior (or a modified likelihood function as technically done in this work). \citet{Burgasser2007} does not use a prior informed by the RV surveys they discuss, though they do note that their posterior distributions have ``non-negligible dispersions, as parameters spaces outside the observational window function (e.g. very tight binaries) add considerable uncertainty". This likely refers to the solutions we saw in our uninformed prior runs and is tentatively seen in the tails of their posterior distributions. \citet{Allen2007} noted a degeneracy between $F$ and $\overline{\log(\rho)}$ which they attribute to resolution limits. We also see this strong degeneracy when using the uninformed prior (and less so when restricting the number of tight companions).  

Since BD mass is degenerate with age for a given flux or luminosity, we ran our fits for a wide range of plausible field ages. \citet{Aganze2022} recently analyzed a sample of ultracool dwarfs out of the galactic plane and modeled the scale-heights, vertical velocity dispersion, and ages. They found disk population ages of $3.6^{+0.8}_{-1.0}$ Gyr for late M dwarfs, $2.1^{+0.9}_{-0.5}$~Gyr for L dwarfs, and $2.4^{+2.4}_{-0.8}$~Gyr for T dwarfs, with an additional 1--2~Gyr systematic uncertainty. With almost 70\% of our sample being L dwarfs ($\sim20\%$ T dwarfs and $\sim10\%$ late M dwarfs) our average field age should be on the lower end of those values, though with such large systematics, the entire range of field ages we used (0.9--3.1~Gyr) are consistent with their measurements. 

Our results show a significantly steeper mass ratio power law index, $\gamma$, than previous studies (depending on the assumed field age). We recover values between $\gamma=4.0^{+1.7}_{-1.5}$ and $\gamma=11^{+4}_{3}$ for assumed field ages of 0.9 and 3.1~Gyr, respectively. Of the three L dwarf studies discussed above, \citet{Allen2007} reports the most shallow index: $\gamma=1.8^{+0.4}_{-0.6}$. They claim their value is softened by the large number of late M dwarfs included in their sample so it is not surprising that our values are larger by $1.4-2.8\sigma$. \citet{Reid2006}, who analyzed a subset of the late M and L dwarfs in this sample, reported $\gamma=3.6\pm1$ which is consistent within $1\sigma$ with our values up to an age of 1.5~Gyr and $2.3\sigma$ lower than our largest value. \citet{Burgasser2007} report a slightly steeper value, $\gamma=4.8^{+1.4}_{-1.6}$, which is consistent within $1\sigma$ with our values up to an age of 1.9~Gyr and $1.8\sigma$ lower than our largest value. \citet{Fontanive2018} report the steepest value of $\gamma=6^{+4}_{-3}$, consistent within $1\sigma$ with all of our values. 

The source of the discrepancy in $\gamma$ values is likely due to our survey's greater sensitivity to lower mass companions near the peak of the separation distribution. Thus, the lack of high contrast/low mass detections can be interpreted as a dearth of such companions rather than a deficiency in sensitivity. Another source of this discrepancy is the difference in how we and previous studies derived our masses/mass-ratios from observed fluxes/flux-ratios. \citet{Reid2006} used 0.5, 1, and 5~Gyr isochrones from \citet{Burrows1997} and \citet{Chabrier2000}, while \citet{Allen2007} used a distribution of ages from \citet{Burrows2001}, and the review by \citet{Burgasser2007} used the original authors' methods,  including the references listed above and dynamical masses. \citet{Fontanive2018} used a distribution of ages and the \citet{Baraffe2003} models. We used a more modern set of models \citep{Phillips2020} which extend to extremely low masses. 

\begin{figure*}
    \centering
    \includegraphics[width=\textwidth]{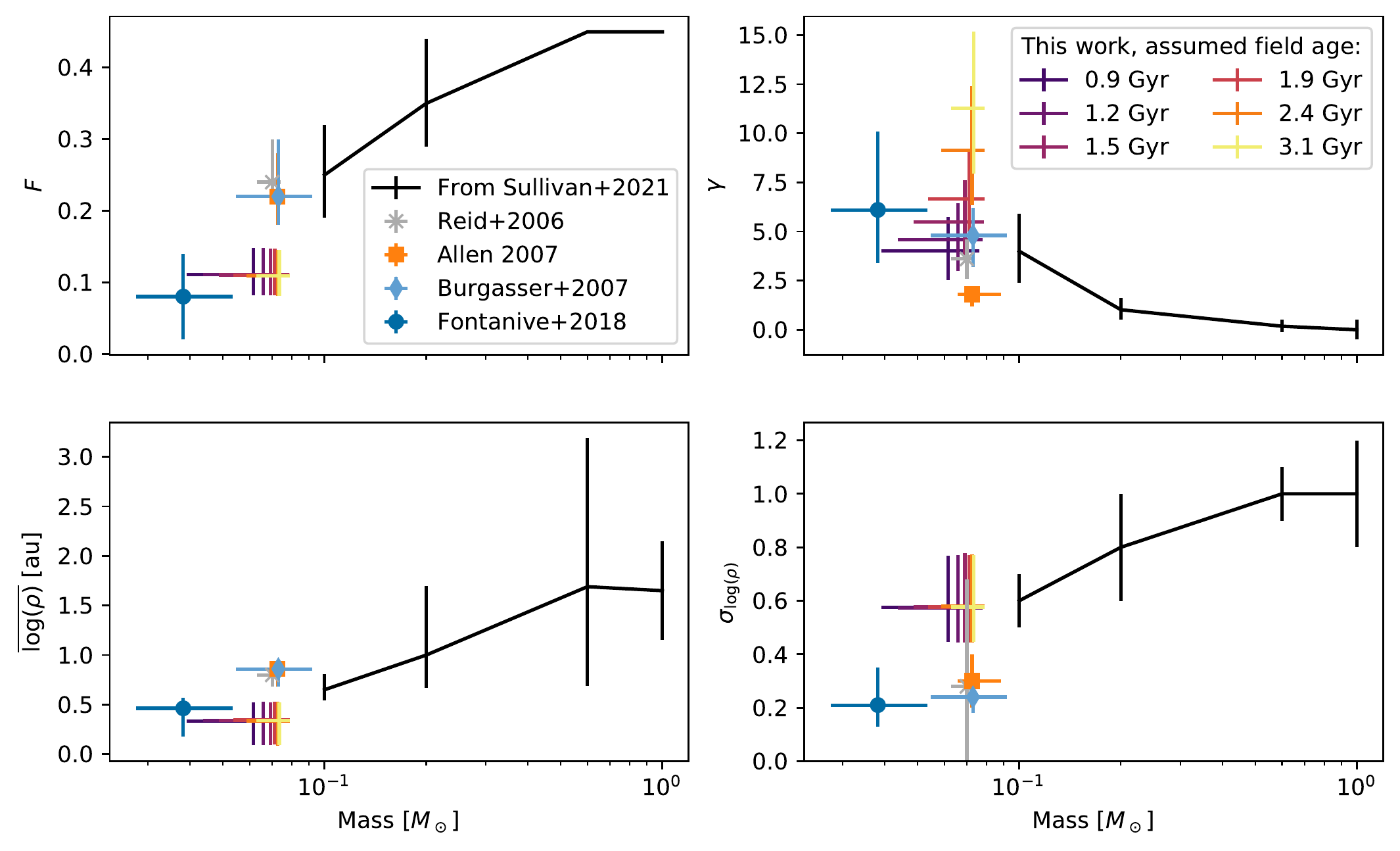}
    \caption{Binary demographic parameters as a function of stellar mass. Black points are taken from Table~1 of \citet{Sullivan2021}, who compiled parameters from a variety of sources \citep{DeRosa2014,Raghavan2010,Kraus2012,Winters2019,Tokovinin2020}. The dark blue circle corresponds to the work of \citet{Fontanive2018} (for T5--Y0), the grey X is from \citet{Reid2006}, the orange square is from \citet{Allen2007}, and the light blue diamond is from \citet{Burgasser2007}. Fits from this work are shown on a color scale from purple to yellow for different assumed field ages. Assumed age mainly affects the derived masses of the sources and $\gamma$, the mass-ratio power-law index, while having little to no effect on the three other parameters. Mass error-bars show the central 68\% interval. We use projected separation, $\rho$, and semimajor axis, $a$, interchangeably as the conversion between the two values is $\sim1$ \citep{Dupuy2011}.}
    \label{fig:paramMass}
\end{figure*}

Figure~\ref{fig:paramMass} shows our best fit parameters in the context of the four BD binarity surveys discussed above and representative values for stellar binaries. Our values follow the general trends as a function of primary mass with discrepancies discussed above. Our study continues to confirm three trends previously noted with BD binarity: a low ($\sim10\%$) binary fraction, a preference for tight pairs (small $\overline{\log(\rho)}$), and a preference for equal mass pairs (large $\gamma$). With our increased sensitivity to close in companions, it appears that the trend in separation could be steeper than previous studies proposed. We also confirm the dearth of wide separation companions seen in previous studies; our posterior binary BD populations have a wide ($>20$~au) observed (corrected for Malmquist bias) companion fraction of $0.9^{+ 1.1}_{-0.6}\%$. We therefore conclude that extremely wide pairs \citep[e.g.][]{Luhman2004,Chauvin2004,Faherty2020} are formed via a different process than most BD binaries or, as is more likely the case, dynamical evolution plays a significant role in BD binary formation as these wide systems tend to also be young.

\begin{figure}
    \centering
    \includegraphics[width=\columnwidth]{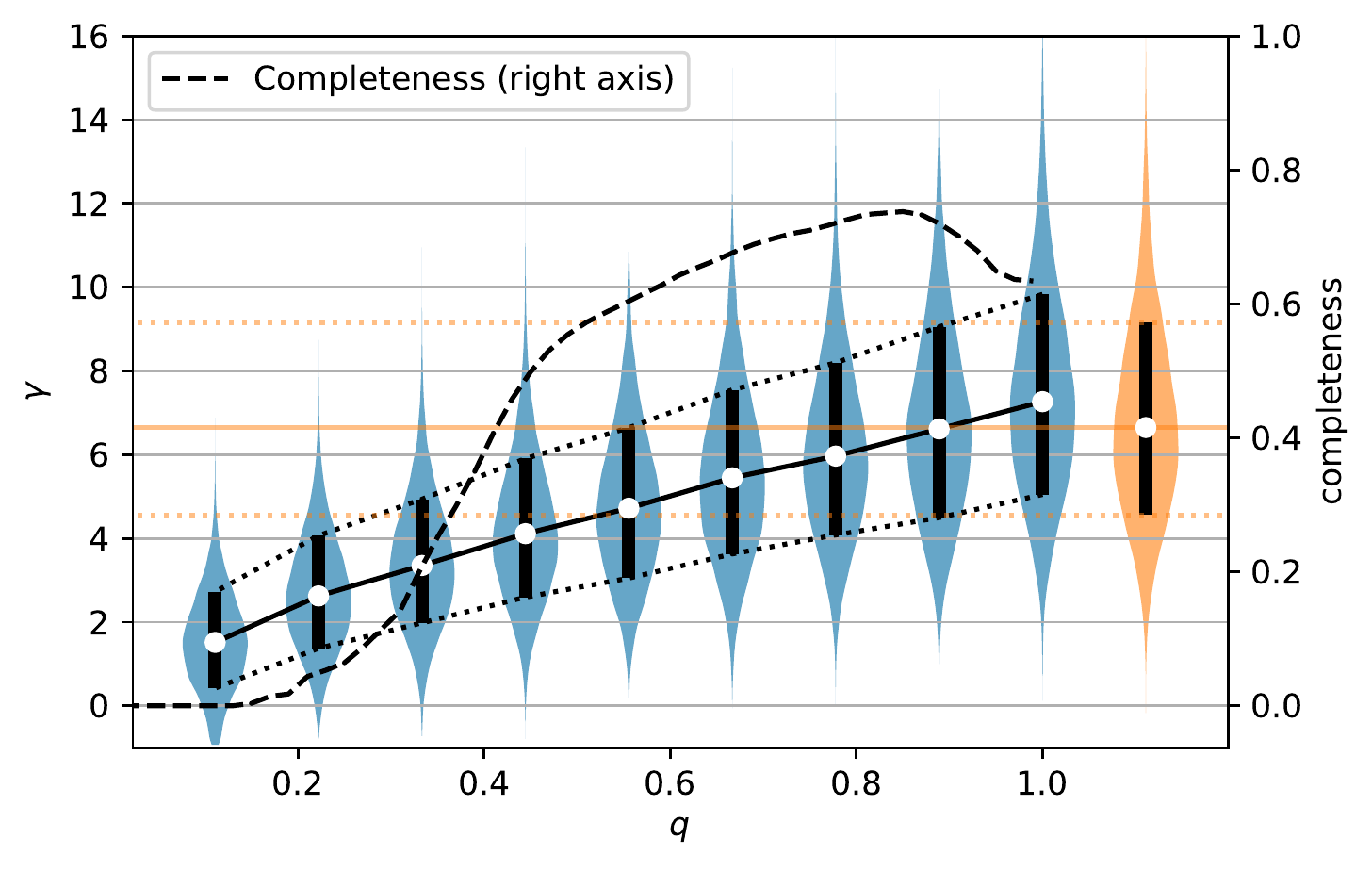}
    \caption{Posteriors of the mass-ratio power-law index, $\gamma$, as a function of the mass-ratio, $q$, of an injected additional binary detection assuming a field age of 1.9~Gyr. The additional detection was injected at the median of the previously derived separation distribution (2.2~au) so as to have the least possible effect on other parameters. The orange distribution is with no injected detection. White points and black vertical lines correspond to the median and $\pm1\sigma$ (central 68\%) values. The black dashed line (right vertical axis) shows the completeness of our survey as a function of $q$. Similar figures for other assumed field ages are available online at \dataset[DOI:10.5281/zenodo.7370349]{https://doi.org/10.5281/zenodo.7370349} \citet{figSets2}.}
    \label{fig:gammaInj}
\end{figure}

Because the mass-ratio distribution is so extreme and the sample size is not huge, it is worthwhile to test the robustness of our results to an additional detection. Figure~\ref{fig:gammaInj} shows the results of refitting our demographic parameters with an additional (synthetic) companion injected at a range of $q$ values (holding separation constant at the median separation). As expected, the inferred $\gamma$ would change only slightly with an additional companion discovered with $q\gtrsim0.6$. While discovering a statistically usable companion (discovered in a broad survey rather than a survey targeting suspected binaries) with a mass ratio less than $q=0.6$ would be rare given the inferred population, it would be extremely significant \citep[e.g. 2MASS 1207b,][]{Chauvin2004} and would change the population parameters significantly. Even assuming our shallowest mass-ratio distribution (a field age of 0.9~Gyr), our posterior binary BD populations have a low mass-ratio ($q<0.6$) companion frequency of $1.0^{+1.4}_{-0.6}\%$ (and even lower for older ages). Similarly to wide companions, we therefore conclude that low-mass companions like 2MASS 1207b are not drawn from the same distribution as most BD binary pairs in the field, again suggesting evolution since 2MASS 1207b is young. 

\subsection{Implications for Binary Formation}\label{sec:discBinForm}

Current state of the art large-scale simulations still do not have the resolution necessary to produce and evolve a statistically significant population of binary BD systems. \citet{Bate2012} calculates a multiplicity fraction of $0.27\pm0.15$ for primaries in the mass range $0.07-0.10~M_\odot$ though the uncertainty is large due to small number statistics. Using the slightly wider mass range of $0.03-0.20~M_\odot$ they calculate a multiplicity fraction of $0.20\pm0.05$, slightly larger than our value by $1.5\sigma$. While \citet{Bate2012} only produced three binary systems with primary masses less than $0.1 M_\odot$, these three systems have a clear bias toward equal masses ($q=0.61,0.94,\mathrm{and~} 0.98$) though only one system had stopped accreting. \citet{Guszejnov2017} also finds a strong preference for equal mass companions for primaries of mass $\sim 0.1M_\odot$. Fitting the data presented in Figure~5 of \citet{Guszejnov2017} we measure a power-law index of $\gamma\sim4.0-4.4$ for their two models. This is consistent at $\sim1\sigma$ with our four smallest (youngest) values and similar to the $\gamma$ value measured by \citet{Burgasser2007}. 

On the other hand, the separation distribution of the three binary BD systems from \citet{Bate2012} seems to be heavily influenced by the resolution of the simulation. The semimajor axis of these systems are 10.6, 26.1, and 36.4~au, two of which are in the rare population of systems with $a>20$~au. Only the 10.6~au system has stopped accreting though, so the other two systems may tighten their orbits with more time. This was also seen in the simulations of \citet{Bate2009}, that BD binaries evolve with time, and observationally by \citet{Close2007} and \citet{Burgasser2007}, that young wide binaries are disrupted by dynamical interactions in the formation environment. \citet{DeFurio2022} studied binary BDs in the Orion Nebula Cluster (ONC) and also found a wide companion fraction significantly above that of the field population, hinting at the importance of dynamical interactions. The simulations presented in \citet{Bate2009} did produce a large number of BD binaries, though again the separation distribution is likely affected by the resolution of the simulation. This distribution had a median separation of 10~au, significantly wider the mean of our separation distribution at $2.2^{+1.2}_{-1.0}$~au. This simulation also uses an older equation of state and no radiative feedback. \citet{Guszejnov2017} does not discuss the semimajor axis distribution for their BD binaries.

Another possible formation pathway for BD binaries is the decay of triple systems \citep{Reipurth2001}. \citet{Umbreit2005} ran an analytical simulation of BD formation integrating the accretion and dynamical evolution of initially triple BD systems. They very accurately reproduce the semimajor axis distribution, with no systems wider than 20~au and a peak at $a\sim3$~au, within $1\sigma$ of our mean separation and significantly smaller that values from previous studies. Fitting a log-normal distribution to the data shown in Figure~8 of \citet{Umbreit2005} we measure a $\sigma_{\log a}\sim0.2$~dex, more consistent with previous studies than with our value. \citet{Umbreit2005} do not discuss the mass ratio distribution or companion frequency of their simulations so it is hard to compare our results in more detail. \citet{Reipurth2015} also conducted a numerical simulation of disintegrating triples which roughly reproduced the observed semimajor axis distribution and strong preference for equal mass companions. However, their simulations produced a significantly higher binary fraction than observed in the field, which they attribute to ignoring the breakup of higher-order multiples. If brown dwarfs form primarily through dynamical interactions such as ejection they should have higher velocities and a larger spatial distribution than stars, which is not seen in young star forming regions \citep{Luhman2012}. This will be interesting to revisit in the Gaia era with higher precision astrometry. 

\citet{Close2003} created a toy model and noted a simple scaling between the mean separation and mass of binaries in the BD and stellar mass regime (i.e. $a_{\circ,\mathrm{BD}}\sim0.13 a_{\circ,\mathrm{TTau}}$ and $M_\mathrm{BD}\sim0.13 M_\mathrm{TTau}$). They argue that this scale factor is set by fragmentation. However, they noted that while mean separation does scale with mass, the width of the distribution does not. Scaling the entire stellar mass separation distribution down to BD mass would produce an extremely large number of wide binaries ($\sim26\%$ of systems with $a>40$~au) which is inconsistent with the observed population. What they do not take into account is the binding energy of these systems. Scaling both the mass and separation does not conserve binding energy since there is a mass squared term. Since the binding energy also decreases with mass, the widest systems would be more likely to be disrupted, thus narrowing the semimajor axis distribution. Assuming binding energy dominates the survival of wide binary systems, scaling the above $\sim26\%$ of systems in ``wide" orbits using binding energy rather than just separation would produce the same fraction of systems wider than $\sim5$~au. Our posterior distribution produces a companion fraction of $3\pm1\%$ for systems wider than 5~au, which is $28^{+11}_{-9}\%$ of systems (from a total companion fraction of $0.11^{+0.04}_{-0.03}$), consistent with this toy model. This also explains the decrease in companion frequency from the stellar mass to BD regime since dynamical interactions tend to disrupt weakly bound systems, rather than hardening them \citep{Kroupa2001,Kroupa2001a,Parker2011}. 

This also helps to explain the preference for equal mass companions in BD binary systems. Stellar mass binaries have a much flatter companion mass distribution, scaling both masses down would maintain this. Since low mass companions have lower binding energy, a low $q$ BD binary is bound more weakly than a similarly scaled stellar mass system, and would be more easily removed thus biasing the mass-ratio distribution toward $q=1$. If this is the case tight binaries should have a flatter mass-ratio distribution than wider binaries. This test requires more detections than we have in our survey and would likely need to consider spectral binaries and directly imaged systems together. 

A process where dynamical interactions play a role naturally implies evolution in the BD binary population. As discussed above, \citet{Burgasser2007} noted that younger BD binary systems have significantly wider separation and flatter mass-ratio distributions, suggesting that wide and low mass companions are initially present from fragmentation but are then removed by field age. Targeted surveys have discovered a number of young benchmark BD binary systems with no direct analogs in the field: Oph 11 \citep{Close2007}, USco CTIO-108 \citep{Bejar2008}, FU Tau \citep{Luhman2009}, 2MASS J0441+2301 \citep{Todorov2010}, Oph 98 \citep{Fontanive2020}. \citet{DeFurio2022} also found a wide ($>20$~au) companion fraction in the young ONC that is significantly higher than the field. 

If dynamical processing does play a significant role in BD binary formation the star-formation environment should have a significant effect on the resulting binary demographics. Many star-forming environments are even denser than the ONC \citep{Lada2003} and would therefore process binary systems more heavily than the ONC. It is possible that much of the field sample came from those environments and would consequently have fewer wide systems. While observing binaries in denser regions than the ONC is difficult due to their distance, looking at the other end of the scale---in the sparsest regions (e.g. Taurus)---is attainable with current archival datasets \citep[e.g.][]{Kraus2005,Kraus2006} and is a prime candidate for reanalysis using higher resolution techniques such as KPI. Currently, tight binaries ($\la10$~au) can only be detected in the field and are not resolvable at the distances of young clusters. Thus, caution must be taken when comparing populations at tight separations as any comparison in this regime is based on extrapolation. For example, using precise PSF template fitting \citet{DeFurio2022a} achieved sensitivity down to $\lambda/D$ or $\sim10$~au at the distance of the ONC and found that a log-normal and power-law separation distribution fit the observed binary M dwarf population equally well. KPI (on the aforementioned observations of nearby star forming regions or on future JWST datasets) could reach separations previously inaccessible to classical imaging techniques and enable longitudinal (i.e. age based) demographic studies at separations approaching the mean of the semimajor axis distribution. 

We thus find it likely that turbulent fragmentation provides the initial conditions for binary BD formation while dynamical evolution modifies the population, dissolving low-mass, wide, and potentially high order multiple systems. This leaves behind a population of tight and equal mass companions as seen in our and other studies. As shown by \citet{Burgasser2003b}, using the framework of \citet{Weinberg1987}, only the widest \citep[$\gtrsim185$~au for BD mass,][]{Close2003} systems are significantly effected by interactions with other field stars, GMCs, and the galactic potential, so this truncation must take place relatively quickly, before the birth cluster dissolves. \citet{DeFurio2022} used the same framework to calculate the lifetime of a wide separation (100~au) BD-BD binary in the ONC to be $\sim5$~Myr ($\sim4-20\times$ shorter than stellar mass binaries) while \citet{Kroupa2003} showed the lifetime of wide ($>20$~au) VLM binaries in a similar environment to be $\sim1$~Myr, both demonstrating the importance of early-time dynamical interactions while the system can still interact with cluster members rather than field stars. Assuming that $\sim10\%$ of BD objects form in low mass clusters (and are thus less disrupted), \citet{Close2007} predict that $\sim0.6\pm0.3\%$ of field BD binaries will be wide ($a>100$~au) which is consistent with their measurement of $f_\mathrm{BD_{wide\&old}}\sim0.3\pm0.1\%$. The question is still open as to why BD binaries seem to be more tightly bound than solar type binaries as noted by \citet{Close2003} and \citet{Burgasser2007}.

\section{Summary}\label{sec:sum}
In this work, we have derived physical properties of the the KPI companion detections and limits from \citet{Factor2022} and fit them with a binary population model. We apply Bayesian modeling techniques adapted from \citet{Allen2007}, \citet{Kraus2011} , and \citet{Kraus2012}, using a Binomial likelihood function to compare our detections to the model distribution. Since we directly consider the detection limits of each observation, this method allows us (and the authors of previous studies) to use all of the observations in a data-set rather than building a volume limited sample and throwing out precious detections. However, we must still account for the Malmquist bias included in our magnitude limited sample which we implement by inflating the number of binaries in our population model according to their contrast ratio. We fit the demographic parameters to our observations using \texttt{emcee} \citep{Foreman-Mackey2013} using wide and flat priors and find more information is required to obtain physically possible results. We thus incorporate limits on the unresolved tight binary population from \citet{Blake2010} and recover values roughly consistent with previous studies though differing in some interesting ways. 

While the overall companion frequency of our underlying population is smaller than previous studies ($F=0.11^{+0.04}_{-0.03}$), the companion frequency of the observed population (without inflating our model to account for Malmquist bias, $F=0.22^{+0.07}_{-0.06}$) is consistent with previous studies. Our separation distribution is closer ($\overline{\rho}=2.2^{+1.2}_{-1.0}$~au) and more broad ($\sigma_{\log(\rho)}=0.58^{+0.20}_{-0.13}$~dex) than previous studies, likely due to the higher resolution of our detection method (KPI) and our incorporation of wide priors with a limit on the unresolved population. Our mass-ratio power-law index ($\gamma=4.0^{+1.7}_{-1.5}$ to $11^{+4}{-3}$ depending on the assumed field age of 0.9 to 3.1~Gyr, respectively) is stronger than previous studies. We attribute this to our different derivation of mass from observed flux and our greater sensitivity to, and non detection of, lower mass (higher contrast) companions. 

\emph{We confirm the trends seen in observational studies over the past two decades of decreasing binary fraction with decreasing mass and a strong preference for tight equal mass systems in the sub-stellar regime.} We attribute this to turbulent fragmentation setting the initial conditions followed by a relatively brief period of dynamical evolution, pruning off the widest and lowest mass companions, before the birth cluster dissolves. Unfortunately large-scale simulations of star formation are still lacking the resolution to produce a large number of binary BDs and thus lack the statistical weight to examine these processes in great detail. We encourage those working on these simulations to keep pushing to higher resolution since these results will provide valuable metrics to compare with observations.   

\acknowledgments
We thank Trent Dupuy and Will Best for useful discussions about this work, and many others who have offered their thoughts at conferences. We also thank the anonymous referee for their helpful feedback which improved the manuscript. This work was funded by \emph{HST} program AR-14561. This work has benefited from The UltracoolSheet \citep[at \url{http://bit.ly/UltracoolSheet},][]{UltraCoolSheet2020}, maintained by Will Best, Trent Dupuy, Michael Liu, Rob Siverd, and Zhoujian Zhang, and developed from compilations by \citet{Dupuy2012}, \citet{Dupuy2013}, \citet{Liu2016}, \citet{Best2018}, and \citet{Best2021}. This work has made use of data from the European Space Agency (ESA) mission {\it Gaia} (\url{https://www.cosmos.esa.int/gaia}), processed by the {\it Gaia} Data Processing and Analysis Consortium (DPAC, \url{https://www.cosmos.esa.int/web/gaia/dpac/consortium}). Funding for the DPAC has been provided by national institutions, in particular the institutions participating in the {\it Gaia} Multilateral Agreement.
\vspace{5mm}
\facility{HST(NICMOS)}


\software{astropy \citep{Astropy2013,Astropy2018},  
          emcee \citep{Foreman-Mackey2013},
          corner \citep{ForemanMackey2016},
          PySynthphot \citep{pysynthphot},
          numpy \citep{numpy},
          SciPy \citep{SciPy}
          }

\bibliography{bib}


\end{document}